\documentclass[usenatbib]{mn2e}
\usepackage{pslatex,graphicx,amssymb,amsmath}
\usepackage{hyperref}
\usepackage{aas_macros}

\title[Time-dependent two-phase black hole accretion disc
models]{Time-dependent models of two-phase
  accretion discs around black holes}

\author[M. Mayer \& J. E. Pringle]{M. Mayer$^1$  \& J. E. Pringle\\
Institute of Astronomy, Madingley Road,
Cambridge CB30HA, UK\\
$^1$ E-Mail: mm@ast.cam.ac.uk}

\begin{document}
\newcommand{\cm}{\textrm{ cm}}
\newcommand{\g}{\textrm{ g}}
\newcommand{\K}{\textrm{ K}}
\newcommand{\jw}{\textrm{j/w}}
\date{\today }

\pagerange{\pageref{firstpage}--\pageref{lastpage}} \pubyear{2006}

\maketitle
\label{firstpage}

\begin{abstract}
We present time-dependent simulations of a two-phase accretion flow
around a black hole. The accretion flow initially is composed of an
optically thick and cool disc close to the midplane, while on top and
below the disc there is a hot and optically thin corona. We consider
several interaction mechanisms as heating of the disc by the corona
and Compton cooling of the corona by the soft photons of the
disc. Mass and energy can be exchanged between the disc and the corona
due to thermal conduction. For the course of this more exploratory
work, we limit ourselves to one particular model for a stellar mass
black hole accreting at a low accretion rate. We confirm earlier both
theoretical and observational results which show that at low accretion
rates the disc close to the black hole cannot survive and is
evaporated. Given the framework of this model, we now can follow
through this phase of disc evaporation time dependently.
\end{abstract}


%\tableofcontents 
\section{Introduction}

X-Ray spectral observations of accretion powered black holes in binary
systems show that there are at least two spectral components
present. The lower energy, thermal component is usually attributed to
the geometrically thin but optically thick disc, which emits as a
black-body or modified black-body \citep[for the latest spectral
model, BHSPEC, see][]{2005ApJ...621..372D,2006ApJS..164..530D}. The
higher energy, power-law component reflects the presence of a possibly
geometrically thick but optically thin, hot corona. The relative
importance of these components changes with time, especially during
periods of strong variability when these sources undergo outbursts. If
the power-law component dominates, the state is called low/hard (LH)
state, while if the black-body dominates the state is called high/soft
(HS) state. The nomenclature 'hard/soft' reflects the shape of the
spectrum, while 'high/low' reflects the presumed accretion rate in
these states. For a review on spectral states, see \citet{0306213}.

In general this picture holds for high-mass black holes, i.e. AGN
(active galactic nuclei), as well. However the low-mass black holes,
i.e. X-Ray Binaries, are the best studied objects with regards to the
temporal behaviour. There the timescale for outbursts is on the order
of weeks to months and years, while in AGN it is much longer. Hence
theoretical models of accretion flows around black holes concerning
the long-term stability properties are observationally best tested
against X-Ray Binaries.

%There are several models of black hole accretion discs available in
%the literature. We briefly introduce a few of them.

Many authors have addressed the thermal and physical states of the
accretion flow in these systems. Much of the work has concentrated on
what is required of the thermal and physical states of the accretion
flow in order to reproduce the X-ray spectra. In these papers, the
system is usually assumed to be in a steady configuration, and the
effort is devoted to computation of the resulting spectrum. There has
also been consideration given to physical and thermal interactions
between the proposed cool (disc) and hot (corona) plasmas, and it is
these considerations which have the strongest bearing on the
time-dependent behaviour of the flow. We briefly discuss some of this
work.

\citet{1975ApJ...199L.153E} and \citet{1976ApJ...204..187S} introduced
the concept of a two-phase accretion model. Their disc consists of an
inner, hot, optically thin and geometrically thick corona, and an
outer, standard, optically thick, but geometrically thin, disc. The
plasma in the inner parts of the corona is a two-temperature plasma,
where electrons and ions interact via Coulomb collisions.  The main
aim of the model was to provide an explanation of the spectrum of Cyg
X-1.

\citet{1991ApJ...367...78W} and \citet{1991ApJ...380...84W} present
models for spectra of Seyfert galaxies given three different
underlying assumptions on the flow geometry. They discuss a
''sandwich'' geometry where the corona and the disc literally form a
sandwich. Corona and disc are either separated radially as in
\citet{1976ApJ...204..187S} or vertically. Models similar to
\citet{1991ApJ...380...84W} were developed by
\citet{1991ApJ...380L..51H,1993ApJ...413..507H}.

While these models were mostly concerned with the explanation of the
observed spectra of the corresponding object, others considered the
structural properties of these two-phase discs and the interactions
between the phases.

\citet{1995MNRAS.277...70Z} and \citet{1997MNRAS.286..848W} present a
two-phase accretion disc model where the two phases interact by
Compton cooling of the corona by soft photons and the disc and heating
of the disc by illumination of the corona, respectively. They carry
out a extensive parameter study of their model, and also include the
effects of mass and energy loss to a wind.

\citet{1990ApJ...358..375B} examine effects of thermal conduction on
two-phase media. They consider the thermal interaction of the hot and
cold phases under isobaric conditions and develop criteria to
determine which of the the hot or cold phase shrinks or grows.

\citet{2000MNRAS.316..473R} discuss radiative and conductive
equilibrium for two-phase accretion disc models. They find
stability/instability strongly dependent on the assumed coronal
heating mechanism. In a later paper \citep{2000A&A...360.1170R} they
consider mass loss/gain of both phases due to thermal conduction for
stationary accretion disc models following the considerations of
\citet{1990ApJ...358..375B}.

\citet{1999A&A...341..936D} presents an semi-analytic model of disc
evaporation by thermal conduction. His model is based on
\citet{1994A&A...288..175M}. He calculates the detailed structure of
the transition layer between disc and corona for an one-temperature
corona. In order to dissolve the disc, he needs a large evaporation
rate which only can be achieved if the conductive length scale is
larger than the radius, so that radial thermal conduction enters the
problem. He gives caveats and questions to what extent this type of
accretion flow can be represented in a 1D model.

One important point in all two-phase models discussed above is how to
determine the relative amounts of luminosity produced in each of the
two phases. Most authors simply introduce a parameter which
predetermines the fraction of luminosity produced or or the fraction
of accretion rate in each phase. \citet{1993PASJ...45..775N} assume
viscous heating to operate in both the hot corona and cool disc. Given
their constraints on the scaleheight of both parts, they find that
most of the heat is produced in the corona.  Some
\citep[e.g.][]{1995MNRAS.277...70Z} constrain this important parameter
through some boundary conditions at the interface between corona and
disc. All of these models however assume a stationary accretion flow.

It was realised fairly early that accretion in a standard optically
thick disc around a black hole is viscously unstable at near-Eddington
accretion rates \citep{1974ApJ...187L...1L}. Then electron scattering
renders the accretion flow unstable.  \citet{1988ApJ...332..646A}
discovered that in these circumstances the radial advection of energy
cannot be neglected. They dubbed such discs ``slim discs''and found
that these instabilities are stabilised by radial advection of
energy. The name comes from the fact that the ratio of disc scale-height to
radial distance, $H/R$, is close to unity (thermal energy close to
kinetic energy), so that the usual 'thin disc' approximation is at
best marginally applicable.  These accretion flows are optically thick
ADAFs (advection dominated accretion flow). These flows are applicable
to the high/soft state.

A similar picture holds for optically thin accretion
flows. \citet{1995ApJ...452..710N} and \citet{1977ApJ...214..840I}
proposed an optically thin counterpart to the optically thick
ADAF. They consider an optically thin disc, i.e. corona, close to the
black hole and an standard optically thick disc further out. Their
model is applicable to lower accretion rates
or the so-called low/hard state. The exact
transition rate from the optically thin ADAF to the optically thick
disc is not predicted by these models, but is determined to match the
observations. 
 
Many of the observed black hole accretion discs show state
transitions, i.e. transitions from the high/soft to the low/hard state
and vice versa \citep{0306213}. The spectrum changes from very soft
and black-body dominated (attributed to the disc) to a hard, power law
dominated spectrum (attributed to the corona). These changes appear to
reflect changes in the accretion rate and flow geometry as discussed
by \citet{1997ApJ...489..865E}. 

In this paper we introduce the various physical concepts which are
required in order to determine the time-dependence of such accretion
flows. Many of the physical properties of such flows are not
well-defined and it is necessary to make many simplifying assumptions
along the way. We shall need to consider three possible disc states,
when the disc is a standard disc with little or no corona, when the
disc has a sizeable corona, and when the disc evaporates altogether
and there is only a corona present. We also need to determine how the
accretion flow transitions from one of these states to another. For
example, if the disc is truncated at low accretion rates and the inner
flow is filled with a hot corona, then we need to determine the
transition radius below which the disc cannot exist any more. Thus we
need to introduce a time-dependent two-phase model for black hole
accretion discs, where mass can be exchanged between the two phases
owing to thermal conduction.

Because the modelling of all these processes is computationally
demanding (apart from all the physical complications this comes about
basically because we need to follow thermal timescales in the inner
hot disc for several viscous timescales of the outer cool disc), we
present here just one example of a time-dependent computation.  We
take as an initial condition a steady-state flow where the hot and
optically thick corona sandwiches a cool and geometrically thin disc
for which the accretion rates in each component are
predetermined. This is similar to some of the models derived in the
literature of the kind invoked to explain the spectral characteristics
of these objects (see above). We then fix the total accretion rate at
the outer boundary and allow the disc to evolve, permitting radial
viscous evolution of both the disc and coronal components, and well as
thermal energy and mass flow between them.

We give a detailed description of the physical assumptions we make to
describe model in Sect.~\ref{sect:model}. These assumptions of
necessity over-simplify the problem and are certainly not unique. they
should however give a reasonable indication of the sort of behaviour
we might expect from such discs. In Sect.~\ref{sect:importance} we
discuss the relative contributions of the corona and disc to the
total luminosity and we also briefly review some observational
results. We then discuss the time-dependent equations for the
two-phase flow in Section~\ref{sect:timedep} and describe the
numerical setup in Sect.~\ref{sect:numerics}. We describe our results
in Sect.~\ref{sect:results}, discuss them in
Section~\ref{sect:discussion} and end with the conclusions in
Sect.~\ref{sect:conclusions}.

\section{The model} \label{sect:model}

We assume that at a given radius the general accretion flow consists
of an optically thick but geometrically thin accretion disc which is
sandwiched by a hot geometrically thick and optically thin corona
above and below. In the following we refer to these parts of the flow
as disc and corona, respectively. Note that at any radius, it may be
that either the disc, or the corona, is not present.

We describe the physics of the disc in Section~\ref{sect:optdisc}, the
corona in Sect.~\ref{sect:corona} and the transition layer in
\ref{sect:tl}. The disc or the corona might cease to exist if either
of them is evaporated, we discuss the treatment of these cases in
Section~\ref{sect:trunc}.

\subsection{The disc}\label{sect:optdisc}

We model the disc as a standard accretion disc in the so-called
one-zone approximation. Hence the density $\rho_\textrm{d}$ and
temperature $T_\textrm{d}$ just represent some average values at a
given radius $R$. Taking constant density, the equation of hydrostatic
equilibrium gives a pressure distribution of the form
\begin{equation}
  \label{eq:hydstatdisc}
  P_\textrm{d}(z)=P_\textrm{d}(0)-\frac{1}{2}\rho_\textrm{d} \Omega_\textrm{d}^2 z^2\;,
\end{equation}
where $\Omega_\textrm{d}$ is the rotational frequency of the disc
which we assume to be Keplerian,
i.e. $\Omega_\textrm{d}=\sqrt{GM/R^3}$. $P_\textrm{d}(z)$ is the pressure at a height $z$ above the midplane. The pressure in the
midplane, which we also take to be the representative pressure for the
disc, contains contributions of gas and radiation pressure,
\begin{equation}
  \label{eq:eosdisc}
  P_\textrm{d}\equiv P_\textrm{d}(0)=\rho_\textrm{d}\frac{kT_\textrm{d}}{\mu m_\textrm{p}}+\frac{4\sigma}{3c}T_\textrm{d}^4\;,
\end{equation}
where we consider a fully ionised, solar metallicity gas. Thus we set the
mean molecular weight $\mu=0.6$. Here $k$ is the Boltzmann constant and
$\sigma$ is the Stefan-Boltzmann constant. 

When the disc is optically thick, the disc radiates as a black body
and we can treat the radiative transfer in the diffusion
approximation, i.e. we write for the one-sided radiative losses in the
disc
\begin{equation}
  \label{eq:qraddisc}
  \Lambda_\textrm{d}^-=\frac{4\sigma}{3\kappa_\textrm{R}\rho_\textrm{d} H_\textrm{d}}T_\textrm{d}^4\;,
\end{equation}
where $\kappa_\textrm{R}$ is the Rosseland mean of the opacity. We use the
tabulated values of the OPAL opacity
project\citep{1992ApJ...401..361R,1996ApJ...464..943I} for the solar
composition of \citet{GN93}. We use the $X=0.7$ set of their opacity
tables ($X$ is the hydrogen mass fraction of the matter) for solar
metallicity ($Z=0.02$).

Energy in the disc is generated by viscosity $\nu$. We use the
parametrisation by \citet{1973A&A....24..337S}, i.e.
\begin{equation}
  \label{eq:viscosity}
  \nu=\alpha c_\textrm{s} H\;,
\end{equation}
where $\alpha$ is the viscosity parameter (typically smaller than
unity, we use $\alpha=0.1$), $c_\textrm{s}\approx \sqrt{P/\rho}$ the
sound speed in the medium and $H$ the scale-height. The viscosity
produces a torque
\begin{equation}
  \label{eq:torquedisc}
  G_\textrm{d}=-\frac{3}{2}\pi\nu_\textrm{d} \rho_\textrm{d} H_\textrm{d} R^2 \Omega_\textrm{d}\;,
\end{equation}
which gives rise to viscous dissipation
\begin{equation}
  \label{eq:viscdisc}
  Q_\textrm{d}^+=\frac{9}{8}\nu_\textrm{d} \rho_\textrm{d} H_\textrm{d} \Omega_\textrm{d}^2\;.
\end{equation}
In both cases we made use of the fact that the disc rotates at the
Keplerian speed, i.e. $\partial (\log \Omega_\textrm{d})/\partial (\log R)=-3/2$.

\subsection{The corona} \label{sect:corona}

As for the disc, we take again density within the corona constant and
equal to some representative value. The pressure distribution in the
corona at height $x = z - H_d$ above the disc/corona interface is
given by
\begin{equation}
  \label{eq:hydstatcorona}
  P_\textrm{c}(x+H_\textrm{d})=P_\textrm{c}(H_\textrm{d})-\rho_\textrm{c}\Omega_\textrm{c}^2 \left(H_\textrm{d}+\frac{1}{2}x\right)x\;,
\end{equation}
where $\rho_\textrm{c}$ is the (representative) density in the
corona, $x$ the height above the disc and $H_\textrm{d}$ the height of
the disc. The point where $P_\textrm{c}(x+H_\textrm{d})=0$ defines the
thickness of the corona, i.e.
\begin{equation}
  \label{eq:hydstatcorona2}
  P_\textrm{c}(H_\textrm{d})-\rho_\textrm{c}\Omega_\textrm{c}^2 \left(H_\textrm{d}+\frac{1}{2}H_\textrm{c}\right)H_\textrm{c}=0\;.
\end{equation}
We set $P_\textrm{c}(H_\textrm{d})=P_\textrm{c}$ to determine the
representative pressure in the corona, and note that $P_\textrm{c} \neq
P_d$. Typically in the corona the representative electron and ion
temperatures ($T_{e,c}$ and $T_{p,c}$) are not equal. The corona is
optically thin, and so we take the representative values of pressure,
density and temperature in the corona to be related by
\begin{equation}
  \label{eq:eoscorona}
  P_\textrm{c}=\rho_\textrm{c}\frac{k\left(T_\textrm{p,c}+T_\textrm{e,c}\right)}{2\mu m_\textrm{p}}\;.
\end{equation}
Note the factor 2 in the denominator of eq.~(\ref{eq:eoscorona}). It
ensures that for a one-temperature plasma
($T_\textrm{p,c}=T_\textrm{e,c}$) the correct value for the gas
pressure is recovered (cf. eq.~\ref{eq:eosdisc}). We assume a pure
hydrogen plasma for the corona (number density of electrons equals
number density of protons). An extension to non-hydrogen plasmas
involves factors of a few percent in~(\ref{eq:eoscorona}). Given the
other approximations made here, not least the simplification of the
one-zone model, we can safely neglect these factors for the purposes
of this paper.
 
We use the same viscosity prescription in the corona as for the disc
(see eq.~\ref{eq:viscosity}), adjusted to the corresponding values in
the corona. The torque in the corona is given by
\begin{equation}
  \label{eq:torquecorona}
  G_\textrm{c}=-\frac{3}{2}\pi\nu_\textrm{c} \rho_\textrm{c} H_\textrm{c} R^2 \Omega_\textrm{c}\;,
\end{equation}
and the heating rate in the corona is
\begin{equation}
  \label{eq:visccorona}
  Q_\textrm{c}^+=\frac{9}{8}\nu_\textrm{c} \rho_\textrm{c} H_\textrm{c} \Omega_\textrm{c}^2\;.
\end{equation}
The scaleheight of the corona may be comparable to the radial distance
from the black hole, i.e. $H_\textrm{c}/R\approx 1$. Then the coronal
flow is pressure supported and deviates from Keplerian rotation.  In
the course of this paper however we keep the assumption of Keplerian
rotation for the corona as well. As long as $H_\textrm{c}/R$ does not
greatly exceed unity, there are only factors of about two
involved. They are however comparable to the likely errors we make in
the vertical averaging and thus we neglect them.

The heating in the corona is distributed onto the electrons and
protons. The fraction of heat going into electrons and protons is not
very well known.  \citet{1998ApJ...501..787G} shows that if the
magnetic fields are near to equipartition, high radiative efficiency
is achieved and most of the viscous heat goes into the electrons,
while for low magnetic fields the heat goes mainly in the ions and low
radiative efficiency, as assumed by ADAF models can be
reproduced. \citet{1998ApJ...500..978Q}, on the other hand, argues that
there is damping at the protons Larmor radius and hence most of the
viscous energy heats the protons. This picture has been reconciled by
\citet{1999ApJ...520..248Q}, who show that for equipartition
magnetic fields most of the energy heats the electrons, while for a
small magnetic field the energy primarily goes into the ions. They
find a transition for $\beta_B=5$ to $\beta_\textrm{B}=100$, where
$\beta_\textrm{B}$ is the ratio of gas to magnetic pressure. The
uncertainty then is introduced by uncertainties how Alfv\'en waves are
converted into whistlers on the protons Larmor radius. In the light of
this discussion, we define our ''fiducial'' heating model for the
corona by heating ions and electrons according to their partial
pressure, $P_\textrm{p,c}$ and $P_\textrm{e,c}$, i.e. we take

\begin{equation}
  \label{eq:qviscci}
  Q_\textrm{p,c}^+=Q_\textrm{c}^+\frac{P_\textrm{p,c}}{P}
\end{equation}
and
\begin{equation}
  \label{eq:qviscce}
  Q_\textrm{e,c}^+=Q_\textrm{c}^+\frac{P_\textrm{e,c}}{P}\;,
\end{equation}
respectively. This, together with the Coulomb collisions, introduced
below, leads to a preferred ion heating for a two-temperature plasma,
and an equal share for a one-temperature plasma.

Electrons and protons are coupled through Coulomb collisions. They
transfer heat from the protons to the electrons at a rate $Q_{ep}$.
We take the rate as calculated by \citet{1983MNRAS.204.1269S}.

The electrons cool radiatively by bremsstrahlung and other cooling
processes (atomic line cooling etc.). While there are cooling rates
for different metallicities and physical conditions available in the
literature \citep[e.g.][]{1993ApJS...88..253S} we limit ourselves here
solely to electron-proton bremsstrahlung. We use the recently
recalculated rate
$\Lambda_\textrm{brems}(T_\textrm{e,c},T_\textrm{p,c})$ by
\citet{2007A&A...461..381M}. The recalculation accommodates the
previously overlooked fact that for a two-temperature plasma the
protons can contribute a significant fraction of the bremsstrahlung
due to their plasma speed. For proton temperatures in excess of
$T_\textrm{p,c}>m_\textrm{p}/m_\textrm{e}T_\textrm{e,c}$ the proton
thermal speed is larger than that of the electron and correspondingly
most of the bremsstrahlung is produced by the deceleration of the
proton rather than the electron. This calculation however strictly is
only valid for the non-relativistic regime, although it could be
generalised to include this if required.

Soft photons from the disc cool/heat the corona by Compton scattering
at a rate \citep[e.g.][]{1979rpa..book.....R}
\begin{equation}
  \label{eq:comptcoolc}
  \Lambda_\textrm{e,c,C}=\Lambda_\textrm{d} \max(\tau_\textrm{es},\tau_\textrm{es}^2) \left[\max\left(4\theta_\textrm{e},16\theta_\textrm{e}^2\right)-\frac{4kT_\textrm{eff,d}}{m_\textrm{e}c^2}\right]\;,
\end{equation}
where $\Lambda_\textrm{d}$ is the soft photon source (see
eq.~\ref{eq:qraddisc}), $\tau_\textrm{es}=\kappa_\textrm{es}\rho_\textrm{c}H_\textrm{c}$ the electron
scattering optical depth ($\kappa_\textrm{es}=0.3\textrm{ cm}^2\textrm{
  g}^{-1}$), $\theta_\textrm{e}=kT_\textrm{e,c}/(m_\textrm{e}c^2)$ and $T_\textrm{eff,d}$ the effective
temperature of the disc. Both an Compton
thick corona ($\tau_\textrm{es}>1$ or slightly relativistic electrons ($\theta_\textrm{e}>1/4$) lead to a strong increase in
the cooling rate and therefore try to keep the electron temperature at
$\theta_\textrm{e}\approx 1/4$, i.e. $T_\textrm{e}\approx 1.5\times
10^{9}$ K, or lower. The first term in the square brackets correspond to
Compton cooling, while the second reflects Compton heating. 
Note that the use of the effective temperature
in~(\ref{eq:comptcoolc}) for Compton heating is valid only if the disc emission
dominates. Then the inverse Compton temperature equals the effective
temperature of the disc. In general, $T_\textrm{eff,d}$ should be
replaced by a term proportional to the flux weighted energy per
photon \citep[e.g.][]{2000ApJ...533..821R}. For the cases presented in
this paper, the electron temperature in the corona is always larger
than $10^8$ K, i.e. Compton heating never plays an important role. 

For the corona we assume a local two-stream radiation field, i.e.
half of the radiation (Bremsstrahlung and Compton radiation) is
directed upwards and leaves
the system, while the other half is directed downwards. If the disc
exists, this part of the radiation field heats the disc at a rate
\begin{equation}
  \label{eq:qheatd}
  Q_\textrm{heat,d}=\frac{1-a}{2}\left(\Lambda_\textrm{e,c,C}+\Lambda_\textrm{brems}(T_\textrm{e,c},T_\textrm{p,c})\right)\;,
\end{equation}
where $a$ is the albedo of the disc. For $a=0$ all radiation is
absorbed while $a=1$ corresponds to a perfect
reflector. \citet{1991ApJ...380L..51H} estimate the albedo to be
between 0.1 and 0.2 by Monte Carlo simulations. Without any
significant loss of accuracy, we set the albedo $a=0$, so that all the
radiation impinging on the disc (if it exists) is assumed to be
absorbed.

\subsection{The transition layer} \label{sect:tl}

We now need to consider the structure within the transition layer
between the disc and the corona. We first assume that the pressure
across the transition zone is constant, i.e. we set
\begin{equation}
  \label{eq:pcpd}
  P = P_\textrm{d}(H_\textrm{d})=P_\textrm{c}(H_\textrm{d})
\end{equation}
and hence in the transition layer of width $\Delta H$ (We essentially
need and assume $\Delta H\ll \left( H_\textrm{d},
H_\textrm{c}\right)$) only density and temperature vary.

In the transition layer heat is exchanged by thermal conduction (see
Sect.~\ref{sect:conduct}). This process leads to mass and energy
exchange between disc and corona by condensation and evaporation. The
amount of mass flowing through the transition layer is assumed to be
sufficiently slow that it does not affect the hydrostatic equilibrium
and sufficiently fast that it can be treated as a stationary,
i.e. time-independent, process, which only depends on the actual
physical state of the corona and disc.

\begin{figure*}
  \centering
  \includegraphics[width=\textwidth]{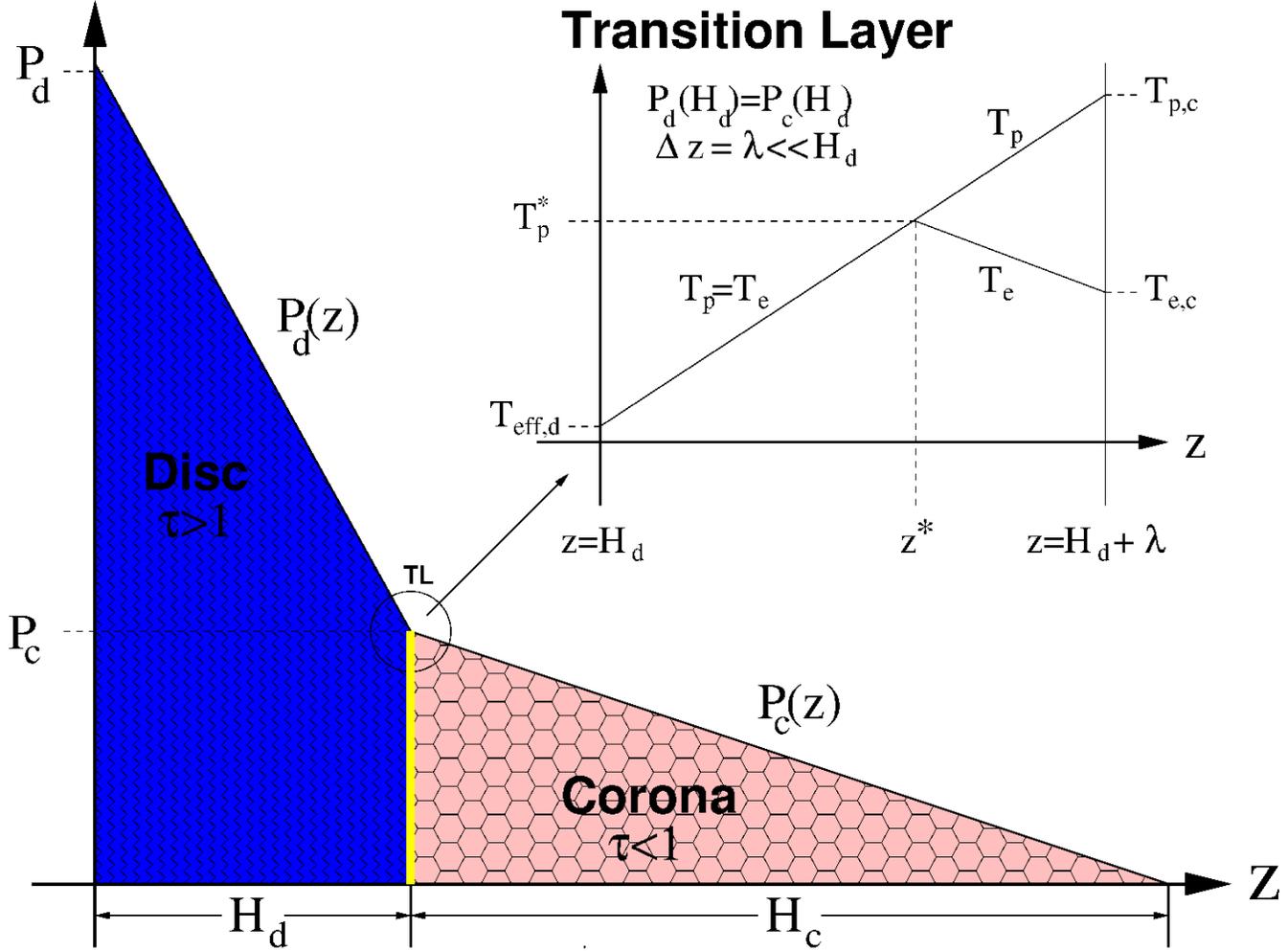}
  \caption{Schematic picture of our disc-corona sandwich, starting
from the midplane (z=0) upwards. The disc has a half-thickness of
$H_\textrm{d}$, while the coronal layer has a thickness of
$H_\textrm{c}$. The average pressure in the disc is approximated by
the pressure in the midplane, $P_\textrm{d}$ and the pressure in the
corona by the pressure at the base of it, $P_\textrm{c}$.  For
simplicity we plot the pressure as a linear function of the height. At
the disc-corona interface the pressure at the base of the corona and
the top of the disc is matched smoothly. We indicate the presumed
temperature structure in the thin transition layer (TL) in the small
inset. The pressure there is constant, but the temperature changes
from the effective temperature of the disc $T_\textrm{eff,d}$ to the
electron and proton temperature of the corona ($T_\textrm{p,c}$ and
$T_{e,c}$), including a (possible) transition to a two-temperature
plasma, if the ion temperature exceeds $T_\textrm{p}^*$. Note that the
length and pressure values are not to scale.}
  \label{fig:schema}
\end{figure*}

We write for the energy transport in the transition layer, following
\citet{1990ApJ...358..375B}
\begin{equation}
  \label{eq:tltrans}
  \frac{d}{dz}\left(\dot m_\textrm{z}\frac{5}{2}\frac{k\left(T_\textrm{p}+T_\textrm{e}\right)}{2\mu m_\textrm{p}}+q_\textrm{cond}\right)=q^+-q^-\;,
\end{equation}
where $\dot m_\textrm{z}$ is the mass flow rate per (disc) surface
area, $q_\textrm{cond}$ is the conductive flux of heat given in
(\ref{eq:qcond}) and $q^+$ and $q^-$ are volume heating and cooling
rates as opposed to the vertically integrated heating and cooling
rates used above. If $\dot m_\textrm{z}>0$, then the disc evaporates
into the corona, and if $\dot m_\textrm{z}<0$ the corona condenses
into the disc.

The characteristic length scale for eq.~(\ref{eq:tltrans}) is the
so-called Field length \citep{1965ApJ...142..531F}, i.e. 
\begin{equation}
  \label{eq:fieldlength}
  \lambda_\textrm{F}=\left(\frac{\kappa_\textrm{cond} T}{\max\left(q^+,q^-\right)}\right)^\frac{1}{2}
\end{equation}

It can be shown that for a one-temperature plasma a stationary corona
(accretion rate independent of radius), where the viscous heating
dominates, leads to a ratio between the Field length and the height
of the corona of
\begin{equation}
  \label{eq:fieldcoronaratio}
  \frac{\lambda_\textrm{F}}{H_\textrm{c}}=\left(\frac{6\pi \kappa_\textrm{cond}(T)T}{\dot M
      \Omega\sqrt{kT/(\mu m_\textrm{p})}}\right)^\frac{1}{2}
\end{equation}

If we assume that the luminosity $L$ of the disc corresponds to the
accretion rate $\dot M$ via $L=\eta \dot M c^2$, where $c$ is the
speed of light and $\eta=1/12$ the efficiency of converting rest mass
into energy, we can scale the accretion rate and luminosity in terms
of the Eddington luminosity and accretion rate $L_\textrm{Edd}$ and
$\dot M_\textrm{Edd}$. We get for the ratio
\begin{equation}
  \label{eq:fieldratio}
  \frac{\lambda_\textrm{F}}{H_\textrm{c}}=1.7\cdot 10^{-3} 
  \left(\frac{R}{R_\textrm{S}}\right)^{\frac{3}{4}} \left(\frac{\dot
      M}{\dot M_\textrm{Edd}}
  \right)^{-\frac{1}{2}} \left(\frac{T_\textrm{e}}{10^9\textrm{ K}}\right)^\frac{3}{2}\left(\frac{T_\textrm{e}}{T_\textrm{p}}\right)^\frac{1}{4}
\end{equation}
Note that the result is independent of the mass. For
$R=300~R_\textrm{S}$ and $T_\textrm{e}=T_\textrm{p}=10^9$ K and $\dot
M=10^{-3} \dot M_\textrm{Edd}$ we get
$\lambda_\textrm{F}/H_\textrm{c}\approx 0.12$. Hence the assumption is
justified, at least with respect to the corona.

The volume heating rate of the disc can be reduced to an expression
only containing pressure and rotation frequency by using the
hydrostatic equilibrium (eq.~\ref{eq:hydstatdisc}) and the dissipation
rate (\ref{eq:viscdisc}) to get
\begin{equation}
  \label{eq:qplusd}
  q^+_\textrm{d}\equiv\frac{Q_\textrm{d}^+}{H_\textrm{d}}=\frac{9}{8}\alpha P_\textrm{d} \Omega\sqrt{2}\;.
\end{equation}
and similarly the dissipation rate of the corona (using
eq.~\ref{eq:hydstatcorona} and \ref{eq:visccorona})
\begin{equation}
  \label{eq:qplusc}
  q^+_\textrm{c}\equiv\frac{Q_\textrm{c}^+}{H_\textrm{c}}=\frac{9}{8}\alpha P_\textrm{c} \Omega\sqrt{\frac{H_\textrm{c}}{H_\textrm{d}+\frac{1}{2}H_\textrm{c}}}\;.
\end{equation}
Note that for a thick corona ($H_\textrm{c}\gg H_\textrm{d}$) the
volume heating rates show the same proportionalities ($\propto \alpha
P\Omega$). For a thin corona on top of a thick disc ($H_\textrm{c}\ll
H_\textrm{d}$), the volume heating rate is reduced by a factor
$(H_\textrm{c}/H_\textrm{d})^\frac{1}{2}$.

The cooling terms can be calculated accordingly. For Compton cooling
we get the volume cooling rate from~(\ref{eq:comptcoolc})
\begin{equation}
  \label{eq:comptcoolctl}
  q^-_\textrm{c,C}\equiv\frac{\Lambda_\textrm{c,e,C}}{H_\textrm{c}}=\kappa_\textrm{es}\rho \Lambda_\textrm{d}\left(\max\left(4\theta_\textrm{e},16\theta_\textrm{e}^2\right)-\frac{4kT_\textrm{eff,d}}{m_\textrm{e}c^2}\right)\;,
\end{equation}
where we assumed a Compton thin transition layer. Given the
geometrical thinness of the layer (see eq.~\ref{eq:fieldratio})
compared to the corona, this approach is justified. The volume cooling
rate~(\ref{eq:comptcoolctl}) is strictly valid only for an entirely
optically thin corona.

We now need to solve (\ref{eq:tltrans}). We follow
\citet{1990ApJ...358..375B} and \citet{2000A&A...360.1170R} and
multiply by $q_\textrm{cond}$ and integrate over the transition layer to get
\begin{equation}
  \label{eq:tltrans2}
  \begin{split}
  \int_{H_\textrm{d}}^{H_\textrm{d}+\lambda_\textrm{F}} &\frac{d}{dz}\left(\dot
    m_\textrm{z}\frac{5}{2}\frac{k\left(T_\textrm{p}+T_\textrm{e}\right)}{2\mu
      m_\textrm{p}}\right)q_\textrm{cond}\; dz \\
  + &
  \frac{1}{2}\left(q_\textrm{cond}(H_\textrm{d}+\lambda_\textrm{F})^2-q_\textrm{cond}(H_\textrm{d})^2\right)=\int_{H_\textrm{d}}^{H_\textrm{d}+\lambda_\textrm{F}} \left(q_\textrm{TL}^+-q_\textrm{TL}^-\right)q_\textrm{cond} \;dz\;.
  \end{split}
\end{equation}
The conductive flux at the boundaries of the transition layer is
zero. If we now define the Field length using the result of
\citet{2000A&A...360.1170R}, then we get
\begin{equation}
  \label{eq:fieldlength2}
  \lambda_\textrm{F}\approx \frac{\int_{H_\textrm{d}}^{H_\textrm{d}+\lambda_\textrm{F}} q_\textrm{cond}
    dz}{\left|\int_{H_\textrm{d}}^{H_\textrm{d}+\lambda_\textrm{F}}\left(q_\textrm{TL}^+-q_\textrm{TL}^-\right)
      q_\textrm{cond} dz \right|^\frac{1}{2}}\;.
\end{equation}
Since the conductive flux contains a temperature gradient, the
integrals over the width of the transition layer in
(\ref{eq:fieldlength2}) can be transformed in an integral over the
corresponding electron and proton temperatures. Note that this
definition of the Field length differs from the one
in~(\ref{eq:fieldlength}), but for the relevant limiting case (coronal
electron and proton temperature much larger than the disc values and
either heating or cooling dominating the other) they are
equivalent. For heating equals cooling the Field length formally goes
to infinity, but then physically there is no mass flux due to
conduction either.

We are only interested in an average mass flux in the transition
layer. Thus we take $\dot m_\textrm{z}=\textrm{const.}$ and hence can
put $\dot m_\textrm{z}$ outside the integral on the right hand side
of~(\ref{eq:tltrans2}). The knowledge of the Field
length~(\ref{eq:fieldlength2}) allows us to estimate an average
conductive flux by setting $\partial T/\partial z\approx
T/\lambda_\textrm{F}$ where we assume that the coronal electron and
proton temperature is much larger than the disc temperature.  We
finally find an estimate for the condensation/evaporation rate $\dot
m_\textrm{z}$ to
\begin{equation}
  \label{eq:evapcond}
\dot
m_\textrm{z}=\frac{\int_{T_\textrm{eff,d}}^{T_\textrm{e,c},T_\textrm{p,c}}
  \left(q_\textrm{TL}^+-q_\textrm{TL}^-\right)
      \kappa_\textrm{cond}
      \left(dT_\textrm{e}+\zeta_\textrm{ep}dT_\textrm{p}\right)}
    {\left|\int_{T_\textrm{eff,d}}^{T_\textrm{e,c},T_\textrm{p,c}}\left(q_\textrm{TL}^+-q_\textrm{TL}^-\right)
      \kappa_\textrm{cond}
      \left(dT_\textrm{e}+\zeta_\textrm{ep}dT_\textrm{p}\right)\right|^\frac{1}{2}}
  \frac{2}{5}\frac{2\mu m_\textrm{p}}{k\left(T_\textrm{e,c}+T_\textrm{p,c}\right)}\;,
\end{equation}
where we set $\zeta_\textrm{ep}=m_\textrm{e}/m_\textrm{p}$.
For the heating we use the viscous volume heating
rate~(\ref{eq:qplusc}), while for the cooling we use the volume rate
for Compton cooling (\ref{eq:comptcoolctl}) and Bremsstrahlung from
$\Lambda_\textrm{brems}(T_\textrm{e,c},T_\textrm{p,c})$, which we
divide by the corona scale-height to get the volume cooling rate,
accordingly.

The sign of $\dot m_\textrm{z}$ crucially depends on the net
effect of the heating and cooling balance in the transition layer. If
the cooling dominates the heating in the transition layer, then hot
matter from the corona condensates into the disc ($\dot m_\textrm{z}<0$). If
the heating dominates, disk matter evaporates into the corona
($\dot m_\textrm{z}>0$).  

We show a schematic picture of the disc-corona sandwich including the
transition layer temperature structure in Fig.~\ref{fig:schema}. In
the inset we show the assumed temperature structure of the transition
layer. The proton temperature is assumed to rise monotonically from
the photosphere, i.e. the effective temperature of the disc, to the
coronal value. At some critical value $T_\textrm{p}^*$ however the
proton electron coupling becomes very weak. Then electrons and protons
start to have different temperatures. We set this critical temperature
to $T_\textrm{p}^*=3\cdot 10^9$ K. For the evaluation of $\dot
m_\textrm{z}$ according to eq.~(\ref{eq:evapcond}) we need to take
this consideration into account. For a one-temperature corona, we
safely can neglect the integration over $T_\textrm{p}$, since it only
enters with a factor of $m_\textrm{e}/m_\textrm{p}$. For a
two-temperature corona, we need to integrate from $T_\textrm{eff,d}$
to $T_\textrm{p}=T_\textrm{e}=T_\textrm{p}^*$ and then until
$T_\textrm{p,c}$ and $T_\textrm{e,c}$, respectively.

We tabulate parts of the integral for selected values of
$T_\textrm{p,c}$, $T_\textrm{e,c}$ and $T_\textrm{eff,d}$ and use
interpolations in these three variables to calculate $\dot
m_\textrm{z}$ in order to save computational time.

\subsection{Truncation}\label{sect:trunc}

At some point in the evolution of the disc-corona sandwich the disc or
the corona can cease to exist at some radius. The determination of the
exact transition to the corona or disc only state however is not
straightforward. We discuss the two cases below.

\subsubsection{Disc truncation}

Here we take the simple approach that whenever the surface density of
the disc falls below a fraction of the  surface density of
the corona at that point, we assume the disc does not exist any
more. We assume the disc ceases to exist if the surface density of the
disc is smaller than 80 per cent of the coronal surface density. The
disc is ''reborn'' if the disc surface density exceeds 90 per cent of
the coronal surface density. 

The corona a priori never becomes opaque to absorption and does not
reach values to become optically thick to electron scattering
either. Hence our disc evaporation criterion approximately coincides
with the point where the electron scattering optical depth of the
disc, $\tau_\textrm{es}=\kappa_\textrm{es}\Sigma_\textrm{c}$ becomes
lower than unity. Then our assumption of optical thick radiative
transfer (eq.~\ref{eq:qraddisc}) cannot be used any more for similar
reasons and different methods need to be used.

If the electron scattering optical depth of the disc becomes lower
than unity, then the high-energy radiation of the corona cannot be
scattered or absorbed either and the heating~(\ref{eq:qheatd}) cannot
operate any more. The disc becomes translucent. To which extent there
can be an optically thin, but cool layer of gas, is not obvious. In
the course of this paper we simply assume that it cannot exist and
these parts of the accretion flow are completely filled with hot
coronal gas.

The truncation criterion used in this paper is somewhat unrealistic,
since even long before the disc becomes optically thin to electron
scattering, it is optically thin to absorption and then it cannot
thermalize internally produced or incoming radiation. The disc will
heat up under such conditions and eventually become part of the
corona. Our assumption however still can be used, since the actual
transition from a disc-corona sandwich to a corona only state takes
place over a very small radial distance. Assuming a more relaxed
condition (i.e. truncate the disc if its surface density is smaller
than $10^2$ times the coronal surface density), the results are still
very similar. 

\begin{figure}
  \centering
  \includegraphics[width=0.5\textwidth]{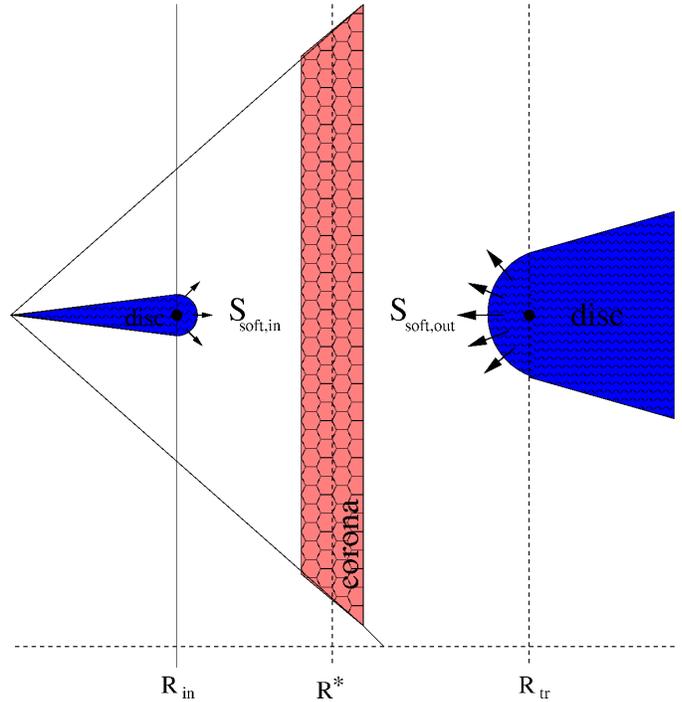}
  \caption{Schematic picture of the truncated disc. The region between
    $R_\textrm{in}$ and $R_\textrm{out}$ is filled with hot coronal
    gas only (a sample coronal column at $R^*$ is shown in pink), while the
    outer and inner parts still consist of an optically thick disc
    with a hot corona above and below. The coronal gas in the
    truncated region can only be Compton cooled by soft photons
    supplied by the outer and inner disc (shown in blue). }
  \label{fig:compton}
\end{figure}

The hot corona-only phase in the middle has to be cooled by soft
photons coming from the disc further out and/or, if it still exists,
from the inner disc as well. We show the situation schematically in
Fig.~\ref{fig:compton}.
The outer disc supplies a total rate of soft photons of
\begin{equation}
  \label{eq:softphotons}
  S_\textrm{soft,out}=4\pi \int_{R_\textrm{tr}}^\infty \sigma
  T_\textrm{eff,d}^4 R dR
\end{equation}
For convenience we integrate to $\infty$, i.e. the outer boundary
of the disc, since $\sigma T_\textrm{eff,d}^4\propto R^{-3}$ and thus
the integral only depends on the value of the effective temperature
close to $R_\textrm{tr}$. At a radius $R^*<R_\textrm{tr}$, only a
fraction of $f=2\pi R^* H(R^*)/(2\pi(R_\textrm{tr}-R^*)^2)$ passes
through the corona, i.e. 
\begin{equation}
  \label{eq:fff}
  f=\frac{R^* H(R^*)}{(R_\textrm{tr}-R^*)^2}\;.
\end{equation}
On the way from $R_\textrm{tr}$ to $R^*$ the amount of soft photons is
reduced due to scattering at radii $R$ with
$R^*<R<R_\textrm{tr}$. Because the corona is optically thin to
electron scattering in vertical direction, the same holds in radial
direction (since $H_\textrm{c}\approx R$). Hence this dilution is not
very significant and we neglect it here. Since $\tau_\textrm{es}<1$,
the dilution factor, $\exp(-\tau_\textrm{es})$ is close to unity. At
$R^*$ the fraction $f$ of the soft photon rate $S_\textrm{soft}$
passes through the area $4\pi R^* H(R^*)$, i.e. we have the soft flux
at $R^*$
\begin{equation}
  \label{eq:softflux}
  F_\textrm{soft,out}=\frac{S_\textrm{soft,out}}{4\pi (R_\textrm{tr}-R^*)^2}
\end{equation}
The electron scattering optical depth in the interval
$\left[R^*-\Delta R/2,R^*+\Delta R/2\right]$ can be approximated by
\begin{equation}
  \label{eq:taues}
  \tau_\textrm{es}(R^*)=\kappa_\textrm{es}\rho_\textrm{c}(R^*)\Delta R\;.
\end{equation}
We calculate the Coulomb cooling rate at $R^*$ by multiplying by
$4kT_\textrm{c,e}/(m_\textrm{e}c^2)$. Division by $\Delta R$ leads to
the corresponding volume cooling rate and multiplication with
$H_\textrm{c}$ to the final cooling rate
\begin{equation}
  \label{eq:coulombcool}
  \Lambda_\textrm{e,c,C,out}=\frac{S_\textrm{soft,out}}{\pi
    (R_\textrm{tr}-R^*)^2}\frac{kT_\textrm{c,e}}{m_\textrm{e}c^2}\kappa_\textrm{es}\rho_\textrm{c} H_\textrm{c}
\end{equation}
Similar arguments hold for the Compton cooling which comes from the
inner disc. Assume the inner discs ends at
$R_\textrm{in}>3R_\textrm{S}$, then the ratio of the Compton cooling
at $R_\textrm{in}<R^*<R_\textrm{tr}$ is
\begin{equation}
  \label{eq:ratiocool}
  \frac{\Lambda_\textrm{e,c,C,in}}{\Lambda_\textrm{e,c,C,out}}=\left(\frac{R_\textrm{tr}-R^*}{R^*-R_\textrm{in}}\right)^2\frac{S_\textrm{soft,in}}{S_\textrm{soft,out}}\;.
\end{equation}
The first factor on the right hand side changes from values much
larger than unity for $R^*\to R_\textrm{in}$ to values much
smaller than unity for $R^*\to R_\textrm{tr}$. The second term, the
ratio between the soft photon rates, is proportional to
$R_\textrm{in}/R_\textrm{tr}$ (Here we assumed again
$T_\textrm{eff,d}\propto R^{-3}$). Hence the contribution of the soft
photon flux from the inner disc is, as expected, only large close to
the inner disc. For the calculation we retain this contribution and
write
\begin{equation}
  \label{eq:coulombcool2}
  \Lambda_\textrm{e,c,C,in}=\frac{S_\textrm{soft,in}}{\pi
    (R^*-R_\textrm{in})^2}\frac{kT_\textrm{c,e}}{m_\textrm{e}c^2}\kappa_\textrm{es}\rho_\textrm{c} H_\textrm{c}\;.
\end{equation}

\subsubsection{Corona truncated}

At some point in the evolution of the accretion flow, the cooling of
the corona can exceed the heating and the scale-height of the
decreases while the density increases. This is especially true for the
parts of the corona further away from the centre. There cooling
mechanisms in addition to those considered in this paper, like atomic
cooling, make the corona thermally unstable and lead to runaway
cooling. The cooling can be halted by Compton heating
(cf. eq.~\ref{eq:comptcoolc} for $T_\textrm{eff,d}>T_{c,e}$). Then the
coronal matter effectively becomes part of the disc. It also may be
halted by the onset of radiation pressure as the optical depth to
absorption approaches or maybe even exceeds unity.  The non-negligible
optical depth will also cause photoionisation heating, since the soft
photons of the disc get absorbed by the collapsing corona. The
examination to which extent photoionisation can stabilise the corona
is likely to be important in some circumstances but is beyond the
scope of this paper. If at some radius the corona completely collapses
and becomes part of the disc, then these parts of the disc are still
heated by the hard radiation coming from nearby disc annuli where the
corona still exists.
 
\section{The relative importance of the corona and the disc for the
  total luminosity} \label{sect:importance}

To give an indication of the disc structure we are likely to be
aiming at, we now estimate what fraction of the luminosity can be
produced by the different parts of a disc and corona sandwich. For
simplicity, to make a rough estimate, we here neglect radial advection
and conductive losses/gains. The results of these estimates give a
strong hint as to the radial extent of the corona and the disc, and we
discuss some of the results from such considerations which are to be
found in the literature.

The stationary, i.e. time-independent version of eq.~(\ref{eq:edisc})
is
\begin{equation}
  \label{eq:ediscstat}
  Q_\textrm{d}^++Q_\textrm{heat,d}=\Lambda_\textrm{d}^-\;,
\end{equation}
and for the corona (combined for electrons and protons)
\begin{equation}
  \label{eq:ecorstat}
  Q_{c}^+=\Lambda_\textrm{brems}+\Lambda_\textrm{e,c,C}^-\;.
\end{equation}

If the disc has an albedo $a$, then the corona heats the disc at a
rate
$Q_\textrm{heat,d}=1/2(1-a)\left(\Lambda_\textrm{brems}+\Lambda_\textrm{e,c,C}^-\right)$.
The remainder of the luminosity created in the corona escapes and
equals the observable luminosity
\begin{equation}
  \label{eq:lc}
  L_\textrm{c}=\frac{1+a}{2}\left(\Lambda_\textrm{brems}+\Lambda_\textrm{e,c,C}^-\right)\;.
\end{equation}
The luminosity of the disc is simply
$L_\textrm{d}=\Lambda_\textrm{d}^-$. The coronal luminosity can be
written as $L_\textrm{c}=(1+a)/(1-a) Q_\textrm{heat,d}$. Hence the
fraction $L_\textrm{c}/L_\textrm{d}$ is
\begin{equation}
  \label{eq:lcld}
  L_\textrm{c}/L_\textrm{d}=(1+a)/(1-a) Q_\textrm{heat,d}/\Lambda_\textrm{d}^-\;.
\end{equation}
It is evident from eq.~(\ref{eq:ediscstat}), that $Q_\textrm{heat,d}$ is always
smaller than $\Lambda_\textrm{d}^-$. Thus the maximum luminosity ratio for a
passively heated disc can be
$L_\textrm{c}/L_\textrm{d}=(1+a)/(1-a)\approx 1.35$ where we set
$a=0.15$ \citep{1991ApJ...380L..51H}. Such a ratio applies for a
passively only heated stationary disc. In all other cases the ratio will be lower.

Coronal luminosities much larger than disc luminosities
(representative for the low/hard state) can be produced in two obvious
situations: either the disc is truncated and the inner accretion flow
solely consists of hot, coronal gas, being cooled by soft photons from
the outer disc, or we view the system along specific lines-of-sight
and take into account relativistic effects.

In the case of X-Ray binaries there is mounting evidence that the disc
is truncated in some states. \citet{1997ApJ...489..865E} gives an
unified picture for black hole accretion flows. They use the accretion
rate as indicator for the different states. For the low-hard and
quiescent states they find a truncated disc, where the inner disc is
filled with an hot coronal advection dominated accretion flow.
\citet{2001ApJS..132..377H} and \citet{2004NuPhS.132..337B} introduce
HID (hardness intensity diagram) as a phenomenological tool to
distinguish different states of X-Ray binaries in a unified
scenario. \citet{2004MNRAS.355.1105F} include contributions from a jet
in this scheme. In their low/hard state they assume a truncated disc.

\citet{2003MNRAS.342..557B} discuss the geometry of the accretion flow
in Cyg X-1. They fit the spectra in the energy range from 3-200 keV
with two models. One model utilises magnetic flares above an
untruncated disc and the other assumes an truncated disc with an inner
hot accretion flow. For the strong reflection in the magnetic flare
case they do not find a break in the spectrum at high energies while
with the weakly illuminated, truncated disc model they easily can fit
the 3-200 keV spectra.

\citet{2004MNRAS.353..980K} discuss the very high-state geometry of
XTE J1550-564. They calculate the luminosity $L_\textrm{d}$ of the
disc emission and show that in the very high state the luminosity is
not compatible with $L_\textrm{d}\propto T_\textrm{d}^4$, where
$T_\textrm{d}$ is the effective temperature of the disc. They conclude
that the simplest explanation might be a smaller effective area where
the disc emission is emitted, i.e. a truncated disc.

\citet{2006MNRAS.367..659D} show that in XTE J1650-500 the broad iron
line often attributed to extreme relativistic smearing can be well
explained in the truncated disc geometry. Then the broad iron line is
formed in an outflowing disc wind and the smearing is significantly
reduced.

For AGN, the situation does not look that dissimilar. However
observable black-body components only can be expected from the very
lower-mass AGN, since only then the effective temperature is
accessible to present X-Ray satellites.

\citet{2004ApJ...606..173P} present observations of NGC 3398. They
fail to find any signs of reflection from an optically thick
disc. Thus they conclude that the disc is truncated at roughly 100-300
Schwarzschild radii.  Similar conclusions are found by
\citet{2005Ap&SS.300...81S} in the case of NGC 7213.
\citet{2006ChJAA...6..165Z} use a truncated disc model to fit
observations of the Seyfert galaxies NGC 5548 and NGC 4051. They find
a disc truncated at 17$\dots$70 and 700 Schwarzschild radii,
respectively.  Models using reflection (therefore assuming an
optically thick disc all the way down to the last stable orbit)
however seem to fit the spectrum equally well \citep[for NGC 4051,
see][]{2006MNRAS.368..903P}.

\section{The time-dependent equations} \label{sect:timedep}

Having introduced the various physical considerations we require, we are
now in a position to introduce the time-dependent equations for our
model.  For the energy equations we follow
\citet{2006MNRAS.368..379M}. We use the mass per unit area (i.e. the
surface density $\Sigma$) and the internal energy $e$ per unit mass in
the disc and corona as independent variables (for the corona we
consider the internal energy for the electrons, $e_\textrm{e,c}$ and
protons, $e_\textrm{p,c}$, separately). All other variables can be
determined by solving eqns.~(\ref{eq:ed}), (\ref{eq:a3}) and~(\ref{eq:a5}) in
Appendix~\ref{sect:trafo}.

\subsection{Continuity equations and angular momentum transport}

For the continuity equations we use
\begin{equation}
  \label{eq:contd}
  \frac{\partial \Sigma_\textrm{d}}{\partial t}+\frac{1}{2\pi
    R}\frac{\partial \dot M_\textrm{d}}{\partial R}=- \dot m_\textrm{z}\;,
\end{equation}
and
\begin{equation}
  \label{eq:contc}
  \frac{\partial \Sigma_\textrm{c}}{\partial t}+\frac{1}{2\pi
    R}\frac{\partial \dot M_\textrm{c}}{\partial R}= \dot m_\textrm{z}\;,
\end{equation}
where $\Sigma_\textrm{d}$ and $\Sigma_\textrm{c}$ are the surface
density and $\dot M_\textrm{d}$ and $\dot M_\textrm{c}$ the accretion
rates in the disc and corona, respectively. $\dot m_\textrm{z}$ is the
mass transfer due to thermal conduction (see
eq.~\ref{eq:evapcond}). $\dot m_\textrm{z}>0$ corresponds to disc
evaporation, while $\dot m_\textrm{z}<0$ corresponds to the
condensation of the corona into the disc.

We calculate the accretion rates from the angular momentum transport
equations where we assume an additional source/sink term for mass
gain/loss. In combination with the continuity
equations~(\ref{eq:contd}) and~(\ref{eq:contc}) this source/sink term
cancels out, since we take/put mass with the same lever arm as
angular momentum is taken/put in \citep[cf. the situation for $R=R_\textrm{A}$
in][]{2006MNRAS.368..379M}. Thus we get for the disc
\begin{equation}
  \label{eq:dmd}
  \dot M_\textrm{d}\frac{\partial \left(R^2\Omega_\textrm{d}\right)}{\partial R}=\frac{\partial G_\textrm{d}}{\partial R}\;,
\end{equation}
and for the corona
\begin{equation}
  \label{eq:dmc}
  \dot M_\textrm{c}\frac{\partial \left(R^2\Omega_\textrm{c}\right)}{\partial R}=\frac{\partial G_\textrm{c}}{\partial R}\;,
\end{equation}
where $\Omega_\textrm{d}$
and $\Omega_\textrm{c}$ are the respective rotation frequencies which
we take to be Keplerian,
i.e. $\Omega_\textrm{d}=\Omega_\textrm{c}=\sqrt{GM/R^3}$.

\subsection{Conductive heat transport} \label{sect:evapenergy}

In Sect.~\ref{sect:tl} we calculate a mass transfer rate owing to the
imbalance of heating and cooling in the transition layer (eq.~\ref{eq:evapcond}). 

If the hot corona condenses into the disc ($\dot m_\textrm{z}<0$), the
residual heat is transferred to the corona at a rate
$S_\textrm{cond}=-\dot m_\textrm{z} e_\textrm{c}$. If the disc
evaporates into the hot corona ($\dot m_\textrm{z}>0$), the corona is
heated at a rate $S_\textrm{cond}=\dot m_\textrm{z}
e_\textrm{d}$. Thus
\begin{equation}
  \label{eq:scond}
  S_\textrm{cond}=
  \begin{cases}
    ~-\dot m_\textrm{z} e_\textrm{c} & \textrm{if $\dot
      m_\textrm{z}<0$ (corona condenses)}, \\
    ~~~~~\dot m_\textrm{z} e_\textrm{d} & \textrm{if $\dot
      m_\textrm{z}>0$ (disc evaporates).}
  \end{cases}
\end{equation}
$e_\textrm{d}$ and $e_\textrm{c}$ are the specific internal energies
per unit mass in the disc and corona (cf. eqns.~\ref{eq:ed} and~\ref{eq:ec}).

\subsection{Energy  equations}

The energy equation for the disc is
\begin{equation}
  \label{eq:edisc}
  \begin{split}
  \frac{\partial }{\partial t}\left(\Sigma_\textrm{d} e_\textrm{d}\right)+P_\textrm{d}\dot H_\textrm{d}=&-\frac{1}{2\pi
    R}\frac{\partial }{\partial R}\left(\dot M_\textrm{d} e_\textrm{d}\right)-\frac{P_\textrm{d}}{2\pi
    R}\frac{\partial}{\partial R}\left(\frac{\dot
      M_\textrm{d}}{\rho_\textrm{d}}\right)\\
  &+\left(Q_\textrm{d}^++Q_\textrm{heat,d}-\Lambda_\textrm{d}^-\right)-S_\textrm{cond}\;,
  \end{split}
\end{equation}
where $\Sigma_\textrm{d}$, $e_\textrm{d}$, $P_\textrm{d}$ and
$H_\textrm{d}$ are the surface density, specific internal energy per
unit mass~(\ref{eq:ed}), Pressure (\ref{eq:eosdisc}) and scale-height
in the disc. $Q_\textrm{d}^+$ and $\Lambda_\textrm{d}^-$ are the local
heating and cooling rates (cf. eq.~\ref{eq:viscdisc} and
\ref{eq:qraddisc}), $Q_\textrm{heat,d}$ the heating by the impinging
radiation of the corona (cf. eq.~\ref{eq:qheatd}) and
$S_\textrm{cond}$ the heat source/sink term for conduction
(\ref{eq:scond}). The first two terms on the right hand side represent
the radial advection of energy and 'pdV'-work, while the others
represent local source/sink terms.

The energy equation for the corona is split into one for the
electrons and one for the protons. We write for the protons
\begin{equation}
  \label{eq:ecorprotons}
  \begin{split}
  \frac{\partial }{\partial t}\left(\Sigma_\textrm{c} e_\textrm{p,c}\right)+P_\textrm{p,c}\dot H_\textrm{c}=&-\frac{1}{2\pi
    R}\frac{\partial }{\partial R}\left(\dot M_\textrm{c} e_\textrm{p,c}\right)-\frac{P_\textrm{p,c}}{2\pi
    R}\frac{\partial}{\partial R}\left(\frac{\dot
      M_\textrm{c}}{\rho_\textrm{c}}\right)\\
  &+\left(Q_\textrm{p,c}^+-Q_\textrm{ep}-\Lambda_\textrm{brems,p}\right)\\
  &+S_\textrm{cond}\frac{e_\textrm{p,c}}{e_\textrm{c}}\;,
  \end{split}
\end{equation}
and 
\begin{equation}
  \label{eq:ecorelectrons}
  \begin{split}
  \frac{\partial }{\partial t}\left(\Sigma_\textrm{c} e_\textrm{e,c}\right)+P_\textrm{e,c}\dot H_\textrm{c}=&-\frac{1}{2\pi
    R}\frac{\partial }{\partial R}\left(\dot M_\textrm{c} e_\textrm{e,c}\right)-\frac{P_\textrm{e,c}}{2\pi
    R}\frac{\partial}{\partial R}\left(\frac{\dot
      M_\textrm{c}}{\rho_\textrm{c}}\right)\\
  &+\left(Q_\textrm{e,c}^++Q_\textrm{ep}-\Lambda_\textrm{brems,e}-\Lambda_\textrm{e,c,C}^-\right)\\
  &+S_\textrm{cond}\frac{e_\textrm{e,c}}{e_\textrm{c}}\;,
  \end{split}
\end{equation}
where $\Sigma_\textrm{c}$, $e_\textrm{p,c}$/$e_\textrm{e,c}$,
$P_\textrm{p,c}$/$P_\textrm{e,c}$ and $H_\textrm{c}$ are the surface
density, specific internal energy per unit mass~(\ref{eq:ec}),
Pressure (\ref{eq:eoscorona}) and scale-height of the corona. The
pressure and internal energy is split into the respective fraction for
protons and electrons. $Q_\textrm{p,c}^+$ and $Q_\textrm{e,c}^+$ are
the heating rates for protons and electrons, respectively (see
eqns.~\ref{eq:visccorona}, \ref{eq:qviscci} and \ref{eq:qviscce}),
$Q_\textrm{ep}$ the electron-proton collision rate and
$\Lambda_\textrm{brems,p}$ and $\Lambda_\textrm{brems,e}$ the
electron-proton bremsstrahlung rate owing to the contributions of
protons and electrons, respectively. $\Lambda_\textrm{e,c,C}^-$ is the
Compton cooling/heating term (eq.~\ref{eq:comptcoolc}). The first two
terms on the right hand side of eq.~(\ref{eq:ecorprotons}) and
(\ref{eq:ecorelectrons}) represent the radial advection of energy and
the 'pdV'-work, while the others are local source/sink terms.

Note that $Q_\textrm{ep}$ is an interaction term and is treated as such in
the energy equations. It conserves energy since the Coulomb collisions
transfer energy from the protons to the electrons. The combined energy
equation for the corona, the sum
of eq.~(\ref{eq:ecorprotons}) and~(\ref{eq:ecorelectrons}) does
reflect this. The conductive energy transport conserves energy as
well as is evident from the sum of all three energy equations. The
same applies for the conductive mass exchange in the continuity
equations.

\section{Numerical Setup} \label{sect:numerics}

Here we discuss the numerical scheme, the initial setup and the
boundary conditions used in this paper.  As we discussed above, the
numerical results presented here are severely limited by the numerical
resources required. We use a logarithmically equidistant grid with
typically $N=30$ points starting from $R=3~R_\textrm{S}$ to
$R=3000~R_\textrm{S}$. All variables except the coronal and disc
accretion rate $\dot M_\textrm{c}$ and $\dot M_\textrm{d}$ are defined
on the grid point. The accretion rate then is defined in between the
grid points as is evident from eq.~(\ref{eq:dmd}) and~(\ref{eq:dmc}).

We take initial values for our simulation from a stationary model
where we neglect radial advection and conduction. The (initial)
accretion rates in the corona and disc are $f_\textrm{c}\dot M$ and
$(1-f_\textrm{c})\dot M$, respectively. 

We integrate the system of time-dependent equations with a one-step
Euler scheme and treat the advection terms in a first-order, upwind
donor cell procedure. We have written our scheme so that we conserve
mass and energy to machine accuracy.

At the inner boundary we impose a zero-torque condition
($G_\textrm{d}(R_\textrm{in})=G_\textrm{c}(R_\textrm{in})=0$). This is
equivalent to setting
$\Sigma_\textrm{c}(R_\textrm{in})=\Sigma_\textrm{d}(R_\textrm{in})=0$. All
matter and energy flowing through this inner boundary is lost from the
system. It is either added to the black hole or lost in a jet. No
advection of energy and mass from within the inner radius is allowed.

At the outer boundary we set the accretion rate $\dot M=0$ for both
the disc and corona. Just inside the outer boundary at the last disc
cell, we supply mass to the optically thick disc at a fixed rate $\dot
M_0=(1-f_\textrm{c})\dot M$. Energy corresponding to the internal
energy of the last disc cell is put into the last disc cell at a rate
$(1-f_\textrm{c})\dot M e_\textrm{d}$. No mass nor energy is supplied to the
corona. Thus $\dot M_0$ is the only parameter characterising the
system. The accretion flow then evolves self-consistently within the
model and mass is transferred between the two phases accordingly.

As one of the computationally more intensive parts of the simulation
is the calculation of the evaporation/condensation
rate~(\ref{eq:evapcond}), we use tables for essential parts of the
integral and use interpolations in the course of the simulations. This
enables us to calculate the evaporation/condensation rate with
sufficient accuracy without spending too much computational time on
the calculation.

The computational time involved in the simulations is
considerable. Since the geometries of the disc and corona are very
dissimilar (the disc is geometrically thin while the corona is
geometrically thick), the viscous timescales are vastly different. The
viscous timescale is \citep[e.g.][]{1981ARA+A..19..137P}
\begin{equation}
  \label{eq:tnu}
\tau_\nu=\frac{1}{\alpha\Omega_K\left(H/R\right)^2}\;,
\end{equation} which in units suitable for the presented models leads
to a value of
\begin{equation}
  \label{eq:tnu2} \tau_\nu=1.4\cdot 10^{-3}\textrm{ s}\cdot
\left(\frac{\alpha}{0.1}\right)^{-1}\left(\frac{M}{10\textrm{
M}_\odot}\right)\left(\frac{R}{R_\textrm{S}}\right)^\frac{3}{2}\left(\frac{H}{R}\right)^{-2}\;.
\end{equation} For the disc usually the ratio between scale-height and
radial distance from the black hole $H/R\approx 10^{-2}$, while for
the corona $H/R\approx 1$.  Thus the ratio of the viscous timescale of
the disc and corona at a given radius is about $10^4$. Thus it takes
$10^4$ times longer for the disc to reach a (possibly) viscous
equilibrium than for the corona. The thermal and hydrostatic timescales,
however are similar.

Numerically the viscous evolution allows for time steps smaller than
\citep{1981MNRAS.194..967B}
\begin{equation}
  \label{eq:dt} \Delta t<\frac{1}{2}\frac{R\Delta R}{12\alpha \Omega_K
H^2}\;.
\end{equation} For $\alpha=0.1$, $H\approx R$ and $\Delta R=\delta R$
with $\delta<1$, it is evident, that the maximum allowed time step for
the viscous evolution is close to or even below the dynamical
timescale.

The dynamical timescale at the outer boundary at $R=3000~R_\textrm{S}$
is about 230 s, i.e. we need to integrate about $2\cdot 10^5$ times
the dynamical timescale of the innermost orbit to reach the dynamical
timescale and 
another $10^4$ times to calculate for one viscous time at the outer
boundary at $R=3000~R_\textrm{S}$ . We then arrive at about $2.3 \cdot
10^6$ s to be calculated in real time, using about $2\cdot 10^9$
time steps. 

Given the different timescales involved in this problem, an implicit
solver would be ideally be suitable.  We however chose to use an
explicit solver since implicit solvers usually have problems with
switching on/off processes (here mainly the mass transfer of
eq.~\ref{eq:evapcond} which switches signs depending on the cooling
and heating in the transition layer) given their large extrapolation
capability. The disadvantage of the explicit solver then is the large
number of time steps needed to evolve the disc.

In order to save computational time, we can make further
approximations. Formally all five time-dependent equations constitute
a coupled system for the time-dependent variables $\Sigma_\textrm{d}$,
$\Sigma_\textrm{d}$, $e_\textrm{d}$, $e_\textrm{p,c}$ and
$e_\textrm{e,c}$
(eqns.~\ref{eq:contd},~\ref{eq:contc},~\ref{eq:edisc},~\ref{eq:ecorprotons}
and~\ref{eq:ecorelectrons}). All the other variables at a given time
can be calculated considering hydrostatic
equilibria~(eq. \ref{eq:hydstatdisc}) and~(\ref{eq:hydstatcorona2})
and the respective expressions for the pressure~(\ref{eq:eosdisc})
and~(\ref{eq:eoscorona}) and energy (\ref{eq:ed} and
\ref{eq:ec}). This system of 5 equations can be reduced to a system of
3 equations (see eq.~\ref{eq:ed} and~\ref{eq:a3}~--~\ref{eq:a5}) but
then it only can be solved numerically. However with a few
simplifications a (computationally expensive) numerical solution can
be avoided. It is evident that since generally $H_\textrm{d}\ll
H_\textrm{c}$, $H_\textrm{d}$ can safely be neglected
in~(\ref{eq:hydstatcorona2}). As we only consider only a disc with
very low accretion rates, we also can neglect radiation for the disc
pressure and internal energy (second term on the right hand sides of
eqns.~\ref{eq:eosdisc} and~\ref{eq:ed}). By applying these
approximations, we can analytically solve the five equations and hence
avoid unnecessary numerical iterations. The same applies to the terms
involving $\dot H$ in both the coronal~(eqns.  ~\ref{eq:ecorprotons}
and~\ref{eq:ecorelectrons}) and disc~(\ref{eq:edisc}) energy
equation. A change $dH$ can be expressed in terms of the surface
density and internal energy of the disc and corona (see
eqns.~\ref{eq:hd} and \ref{eq:hc}). In the course of this paper, we
choose to neglect this term in order to save computational
time. Physically this is justified since the corona and disc only
influence each other through this term if the pressure in the disc and
corona are close to each other or the scale-height of the disc and
corona are similar (cf. eqns.~\ref{eq:hd} and~\ref{eq:hc}). Then
however the disc or corona ceases to exist and other measures need to
be taken (cf. Sect.~\ref{sect:trunc}). In all cases other than that
the coupling involving the $\dot H$ terms is negligible.

\subsection{Disc develops a hole} \label{sect:truncrec1}

Our recipe for the truncated part of the disc are as follows. If the
surface density of the disc falls below 80 per cent of the surface
density of the corona, the disc does not exist any more. The
corresponding cell however is allowed to advect mass and energy from
the neighbouring still active disc cell.  The disc is ``reborn'', if
the ``dead'' disc surface density has refilled up to 90 per cent of
the coronal surface density.  ``Dead'' disc cells do not have thermal
conduction, produce no torque, are neither heated by the corona nor by
viscous dissipation. The corona in this parts of the flow is cooled by
soft photons from the outer and inner (as long as it exists) disc
parts (cf. Sect.~\ref{sect:trunc}).  The innermost cell of the outer
disc is defined as the cell where most of the mass is evaporated in
the corona. The outermost cell of the inner remnant disc still loses
mass to the next ``dead cell'' for angular momentum conservation. If
the surface density, i.e. mass, of the disc in this cell is
sufficiently high, then the disc is reactivated again. As the surface
area of the disc annuli increases proportionally to $R^2$, the
outwards advected mass hardly ever manages to reactivate a disc cell.

\subsection{Corona fully condenses into disc}\label{sect:truncrec2}

This case does not occur in the model presented in this paper. If the
corona did fully condense into the disc, then the corresponding
torques and accretion rates in this parts of the corona decrease
towards zero. If the scale-height of the corona shrinks below some
limit (Physically we should take the point when the optical depth with
respect to absorption becomes larger than unity), this part of the
flow then is in a disc only state. Mass and energy influx from either
side into the corona is still allowed. The corona starts to exist
again if the optical absorption depth of the corona has shrunk below a
certain threshold.

\begin{figure*}
  \centering

  \begin{tabular}{cc}
    \includegraphics[width=0.5\textwidth]{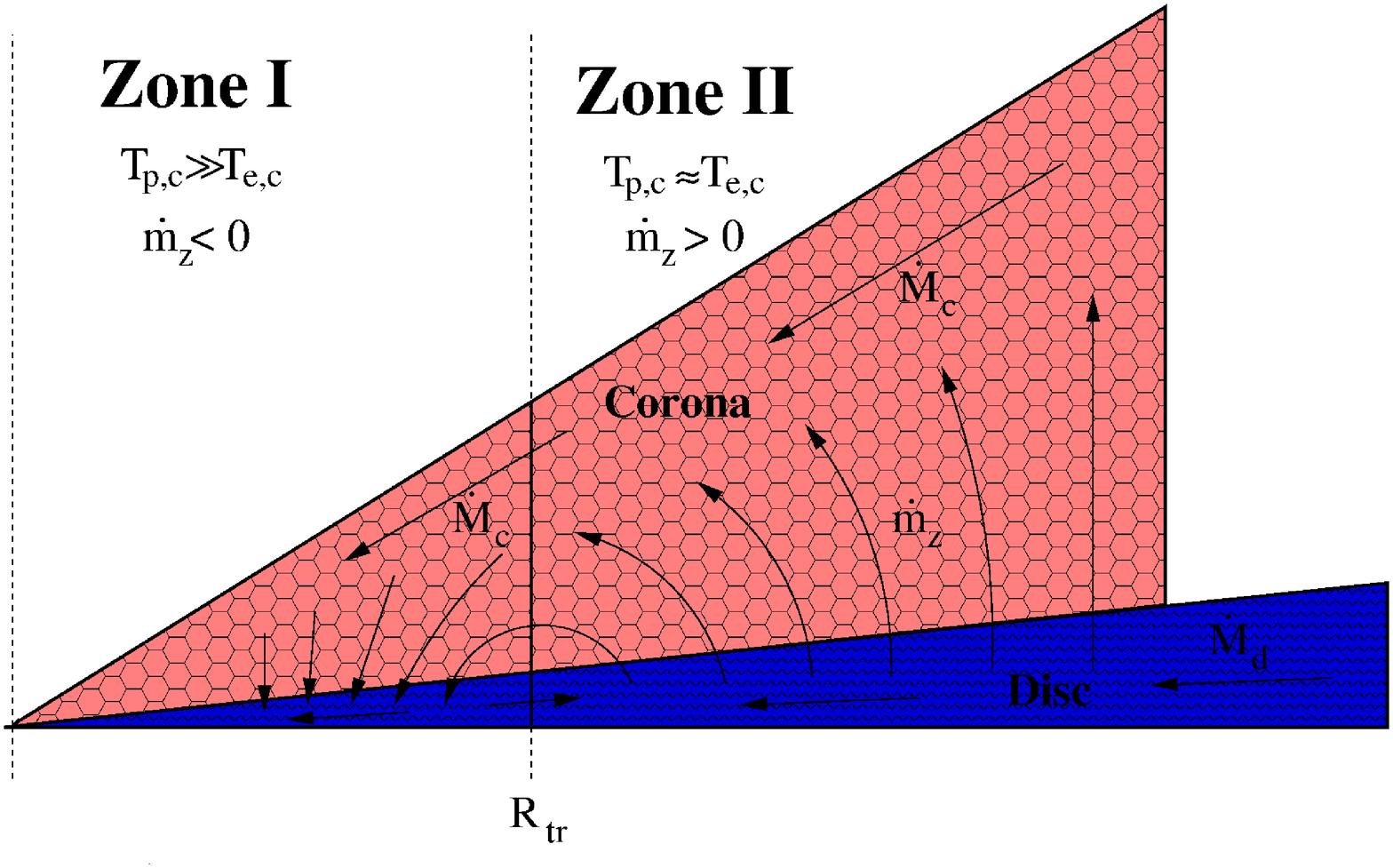} & {\Large \bf 1} \\
    \includegraphics[width=0.5\textwidth]{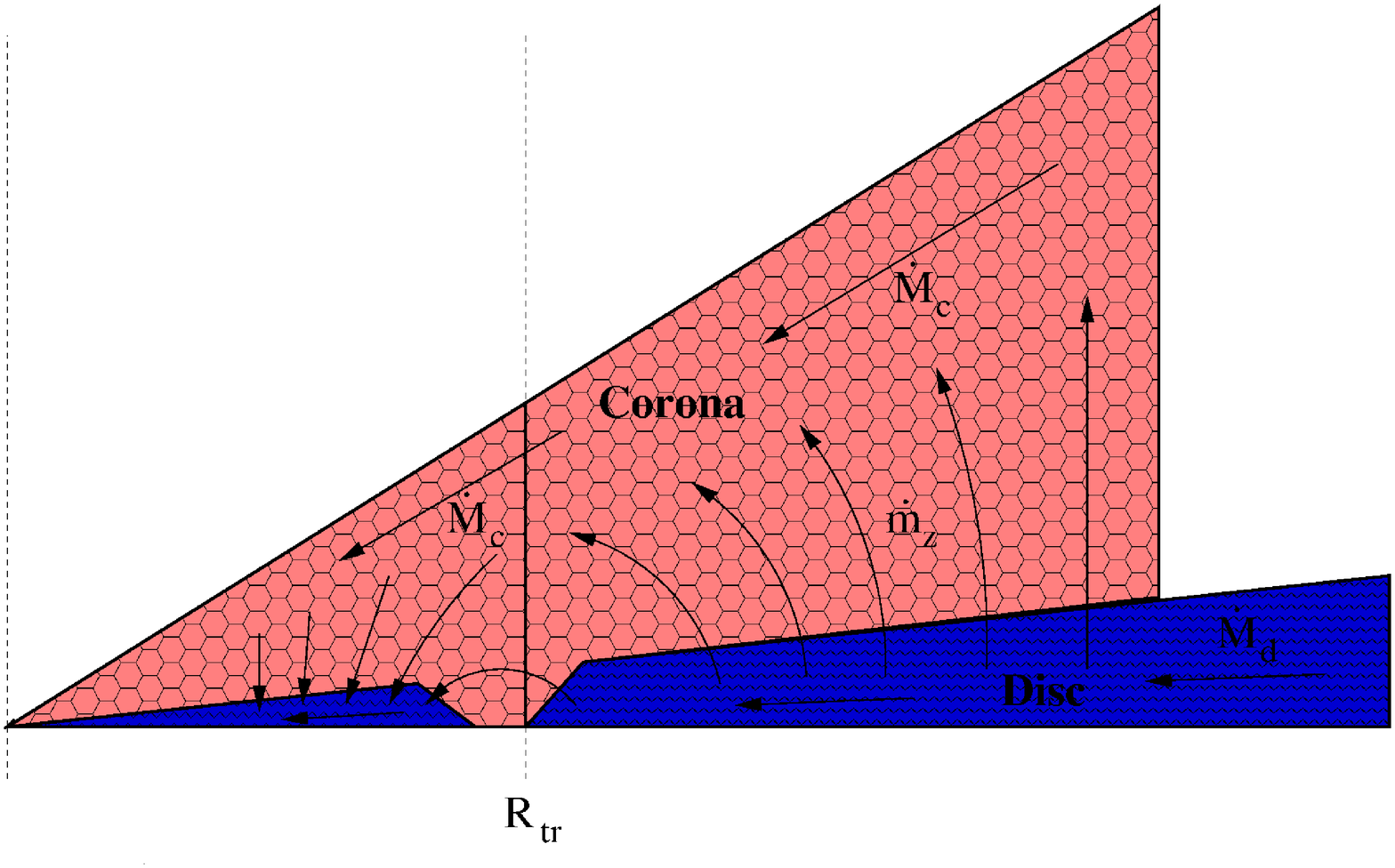} & {\Large \bf 2} \\
    \includegraphics[width=0.5\textwidth]{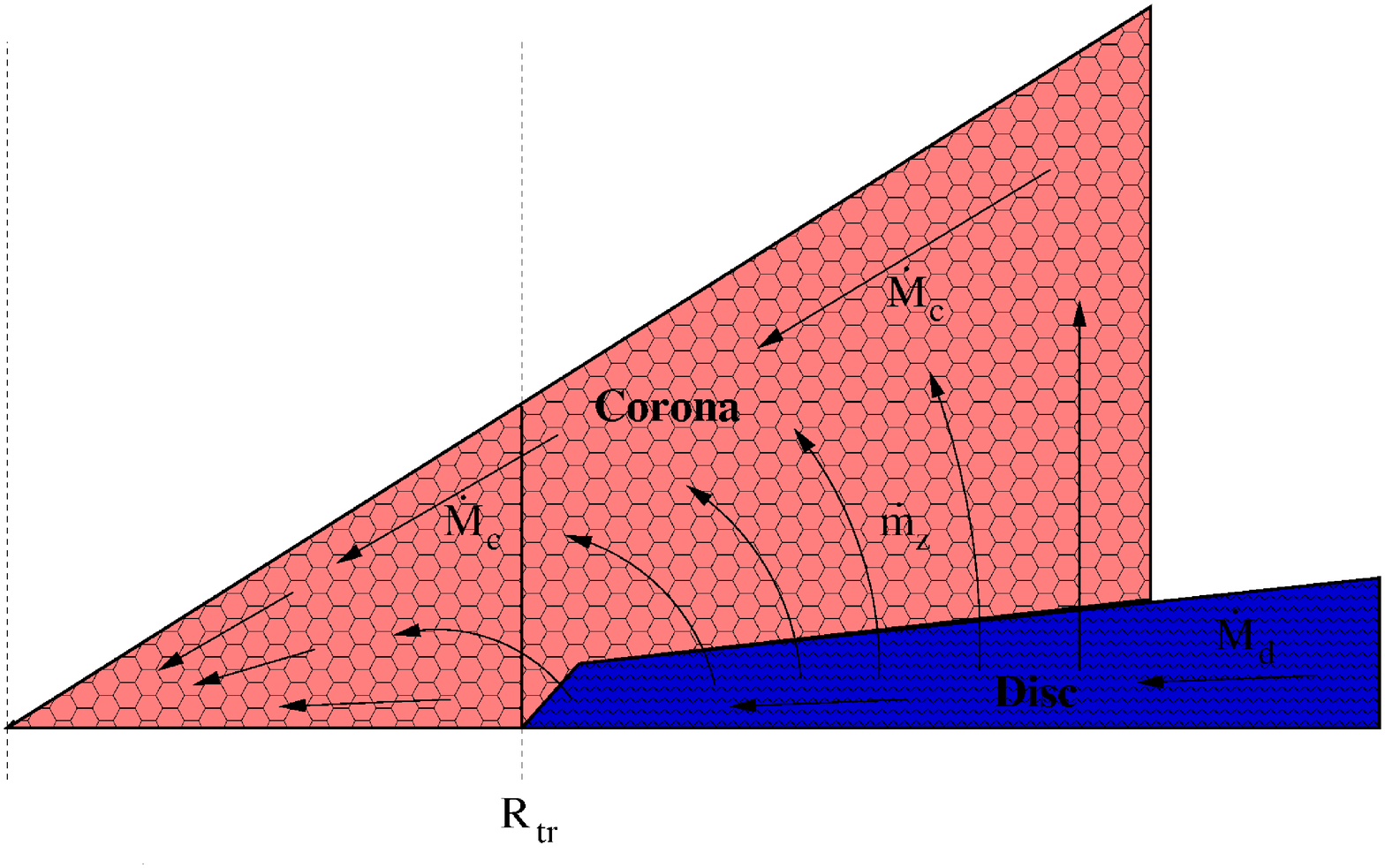} & {\Large \bf 3}
  \end{tabular}

  \caption{Schematic view of a typical
sandwich disc in the course of the time-evolution. We only show one
side of the sandwich. The sandwich is symmetric with respect to the
horizontal axis. The arrows indicate
the mass flow within the corona (light red) and within the disc (blue)
as well as the
conductive mass flow between the two phases. The flow radially
comprises of three zones: In Zone II (one-temperature
corona) the disc evaporates leading to the formation of a corona. The
evaporated mass flows inwards within the
corona. In Zone I (two-temperature corona) the corona
condenses into the disc. Both the disc and the corona accrete in this
zone. At about the transition radius $R_{tr}$ disc flows outwards
(Panel 1). 
After some time, the disc is truncated and the hole is filled with
coronal gas (Panel 2). The inner disc quickly gets accreted and
the system is left with the situation as shown in Panel 3.}
  \label{fig:schemat}

\end{figure*}

\section{Results} \label{sect:results}

In the calculation we describe here, we consider an accretion disc
model for a 10 M$_\odot$ black hole accreting at $10^{-3}$ $\dot
M_\textrm{Edd}$. We describe the time-evolution below.

\begin{figure*}
  \centering
  \begin{tabular}{cc}
    \includegraphics[width=0.48\textwidth]{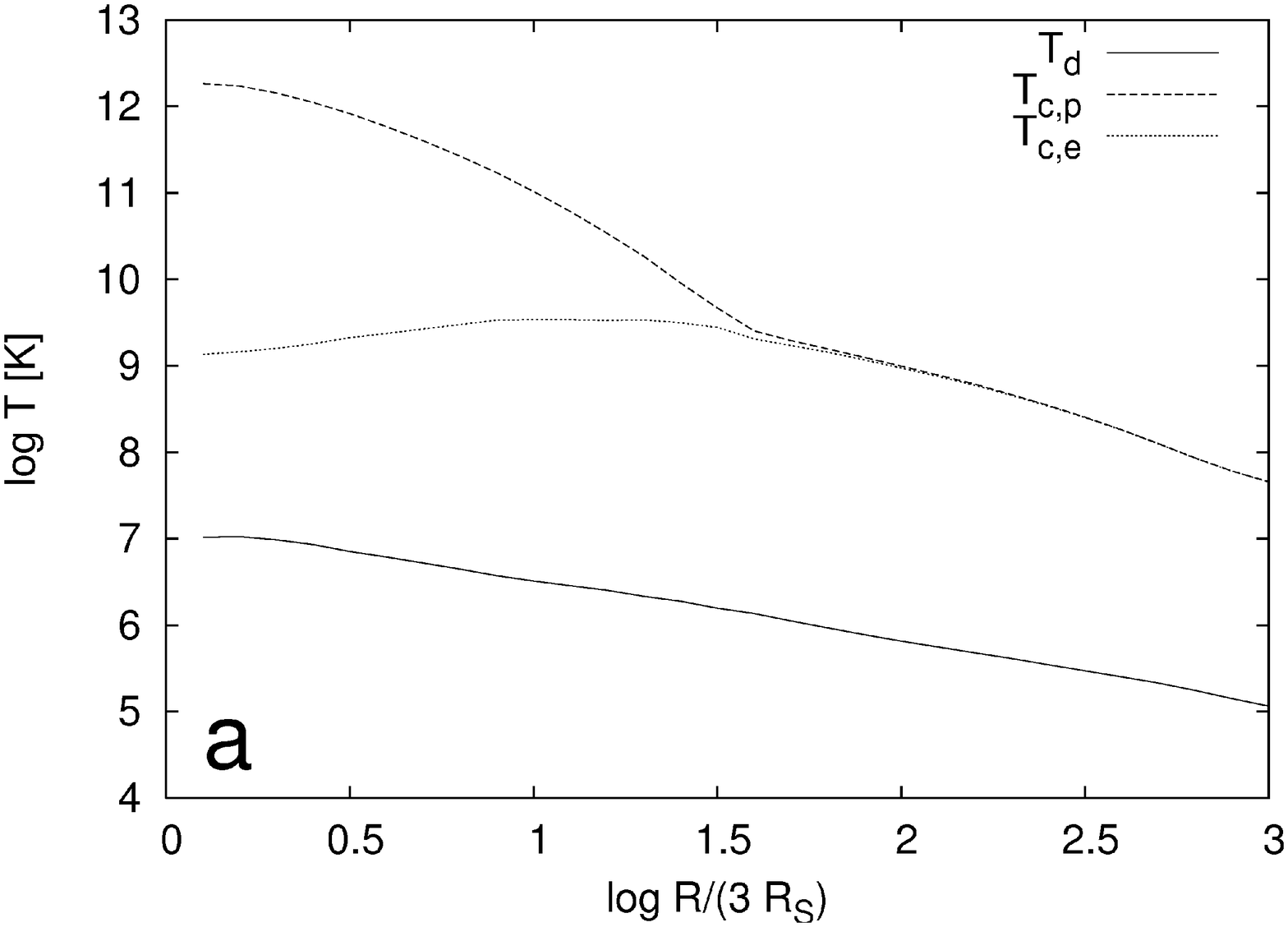}
    &\includegraphics[width=0.48\textwidth]{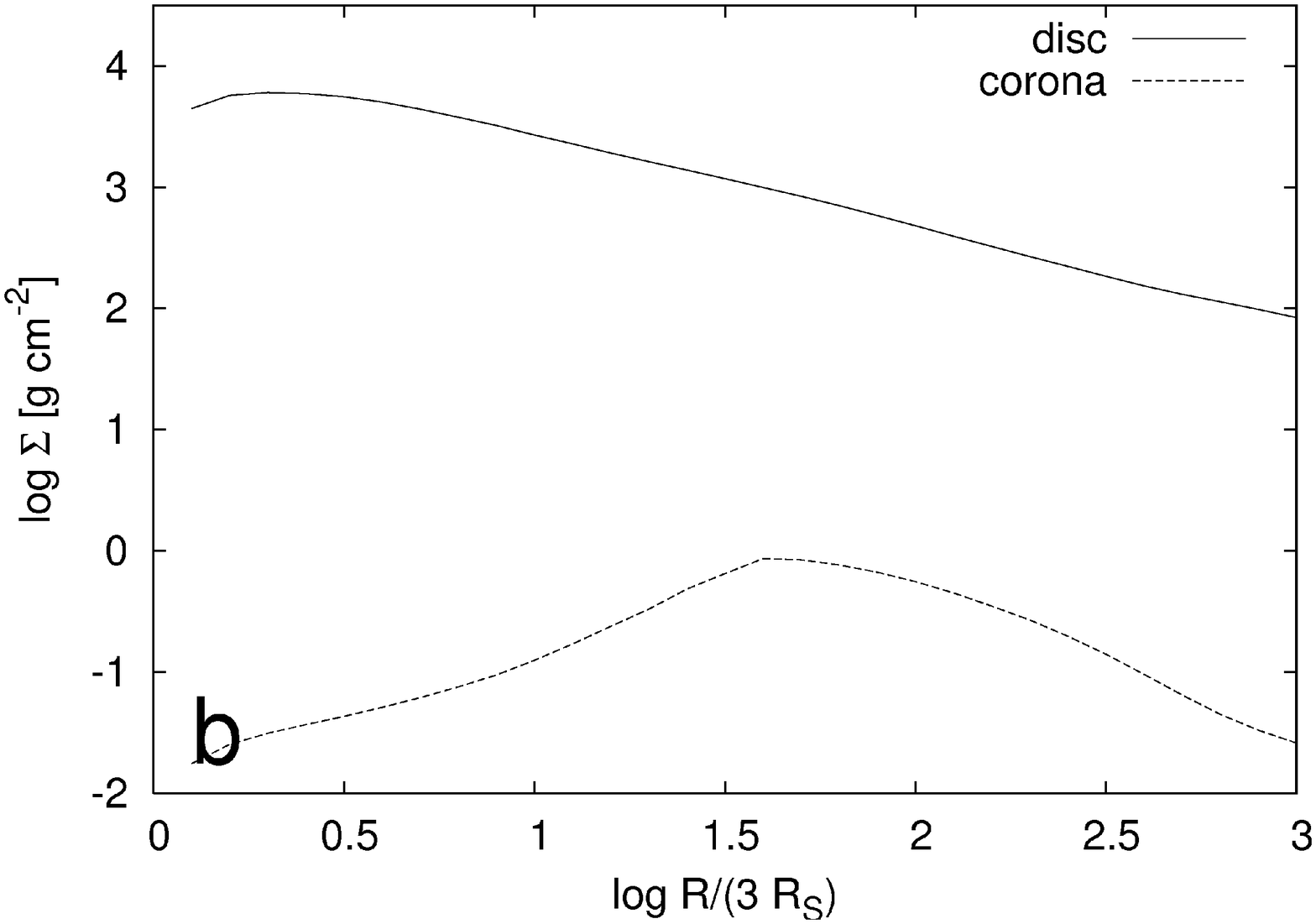}\\
    \includegraphics[width=0.48\textwidth]{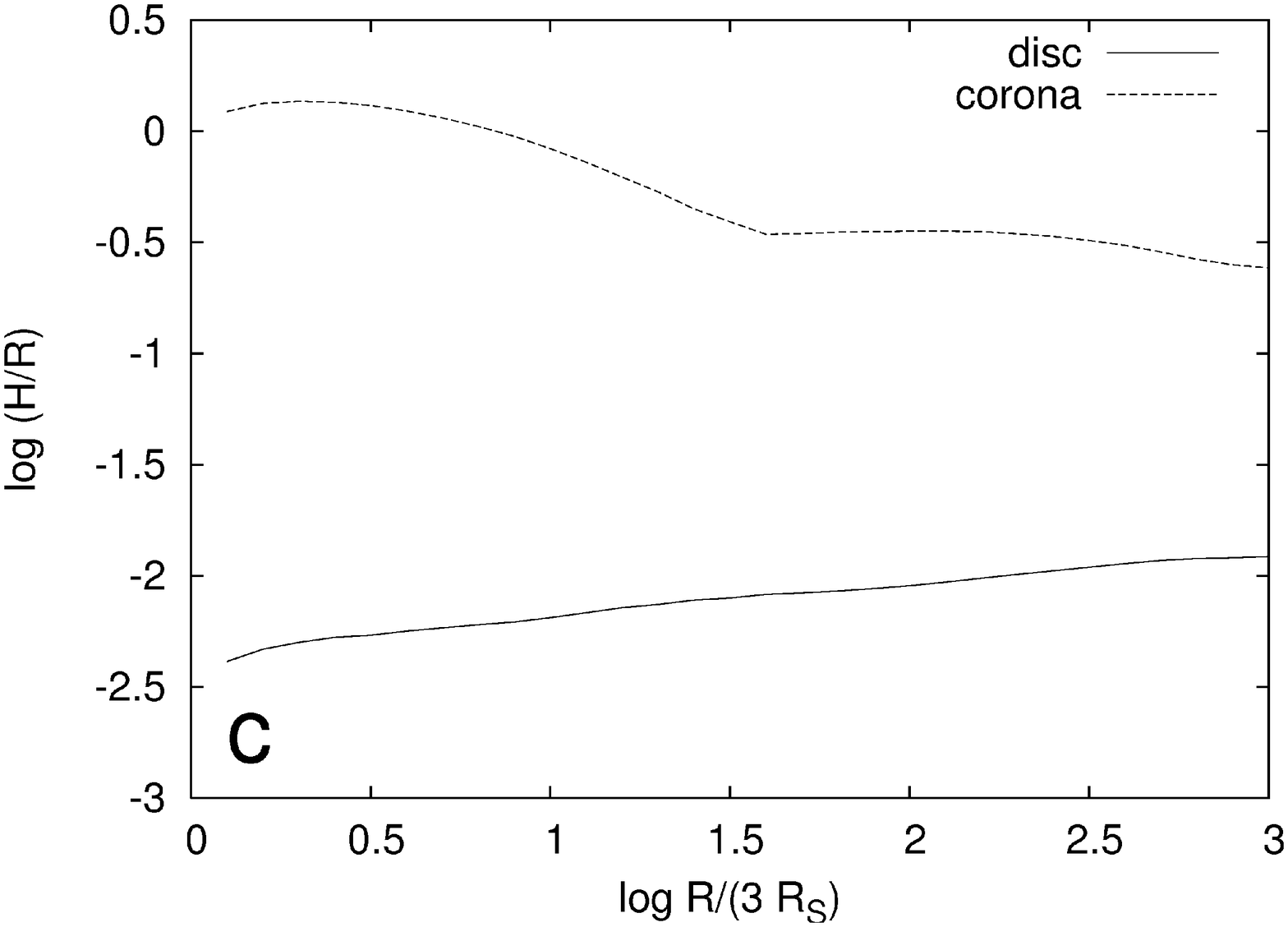} 
    &\includegraphics[width=0.48\textwidth]{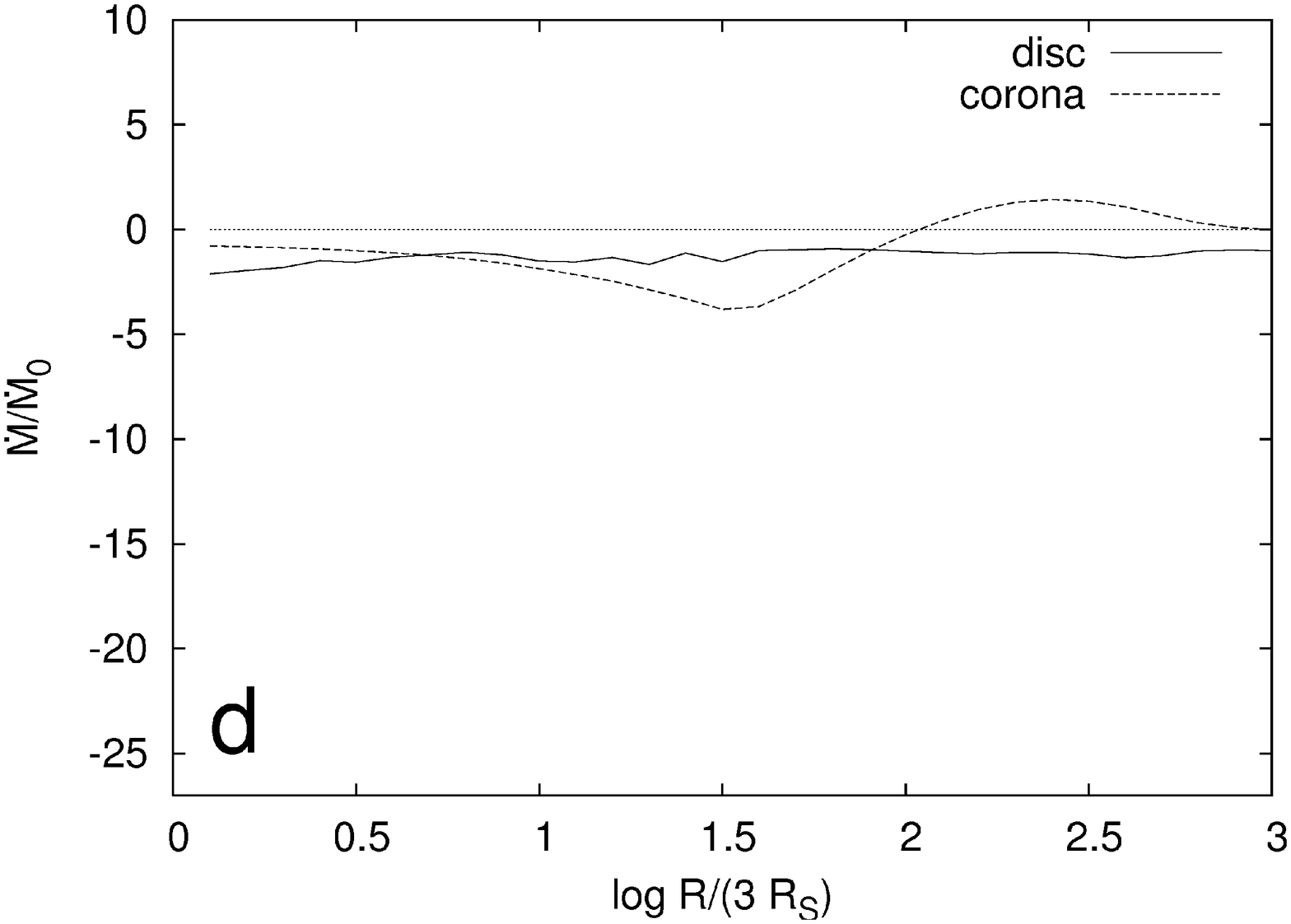}
  \end{tabular}
  \caption{Some physical quantities in the
    time-evolution of the accretion flow 100 s after the start of the
    simulation: Panel a shows the electron and proton temperature in
    the corona and the temperature of the disc, Panel b,c, and d show
    the surface density, the ratio (H/R) and the accretion rate for
    disc and corona, respectively. Note that we normalized the
    accretion rate to the rate we feed the disc in the outermost grid
    cell. The disc is evaporating in its
    outer parts very strongly and leads to an outflowing corona beyond
  a critical radius (see Panel d). }
  \label{fig:schemap1}
\end{figure*}

\begin{figure*}
  \centering
  \begin{tabular}{cc}
    \includegraphics[width=0.48\textwidth]{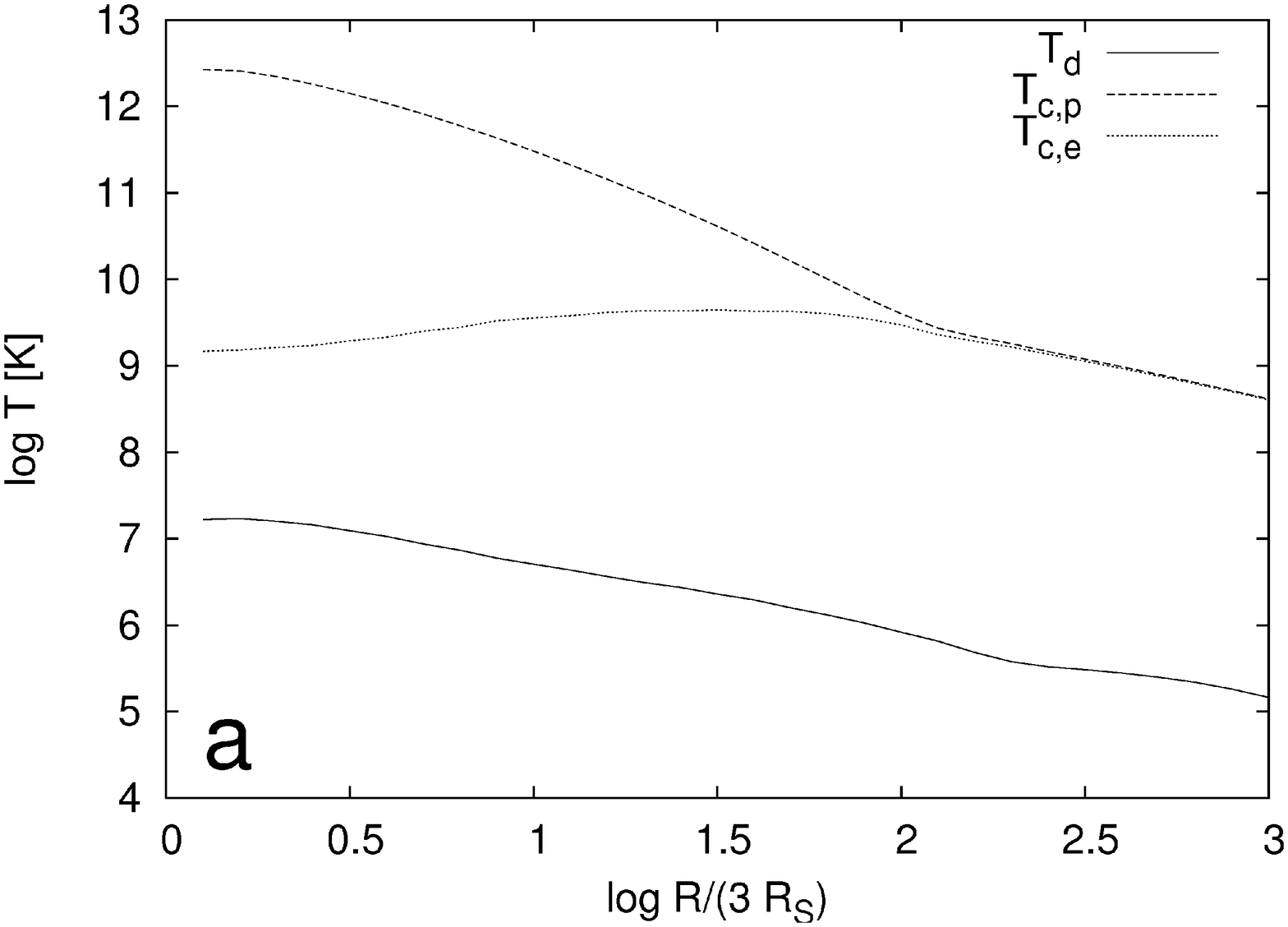}
    &\includegraphics[width=0.48\textwidth]{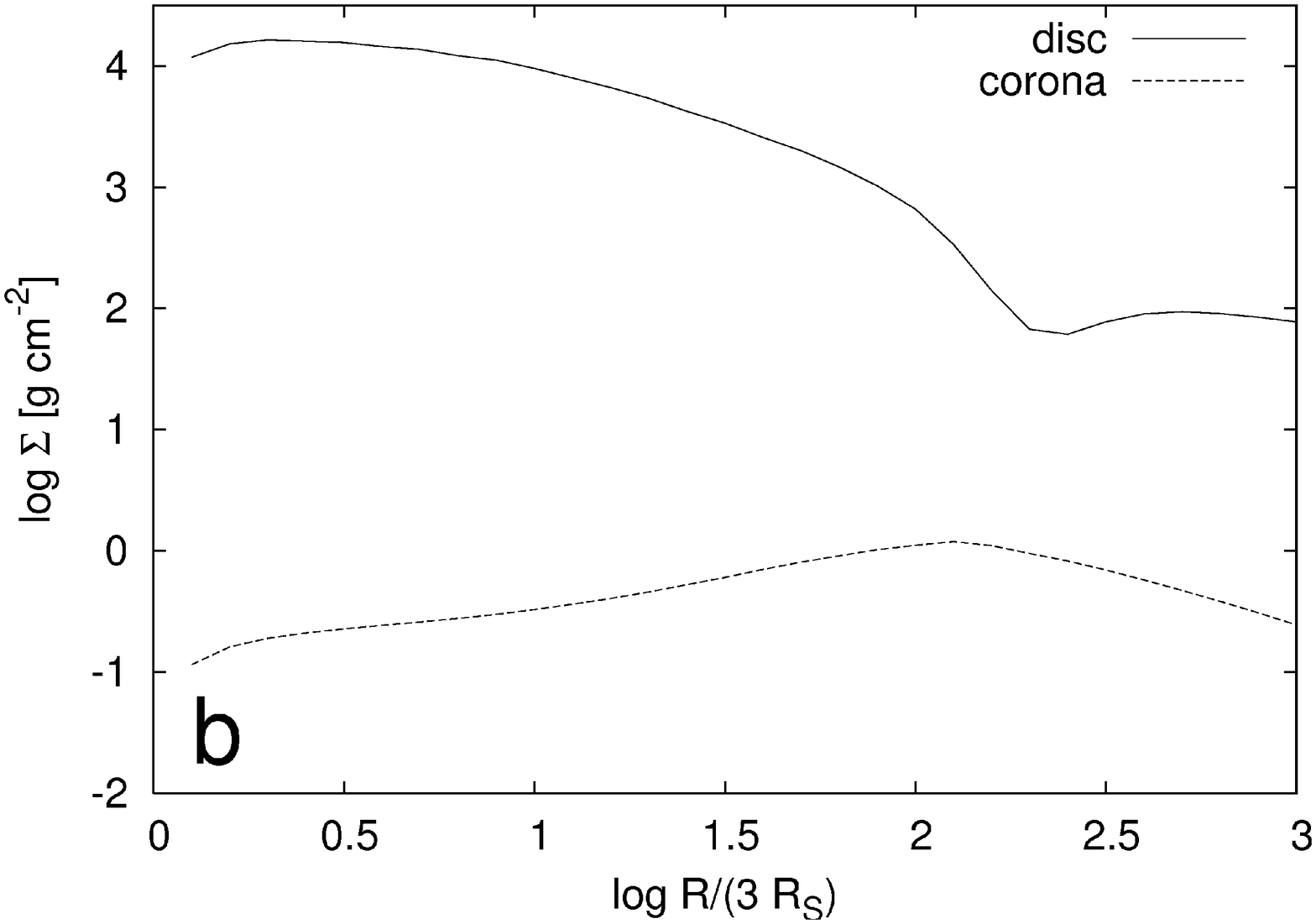}\\
    \includegraphics[width=0.48\textwidth]{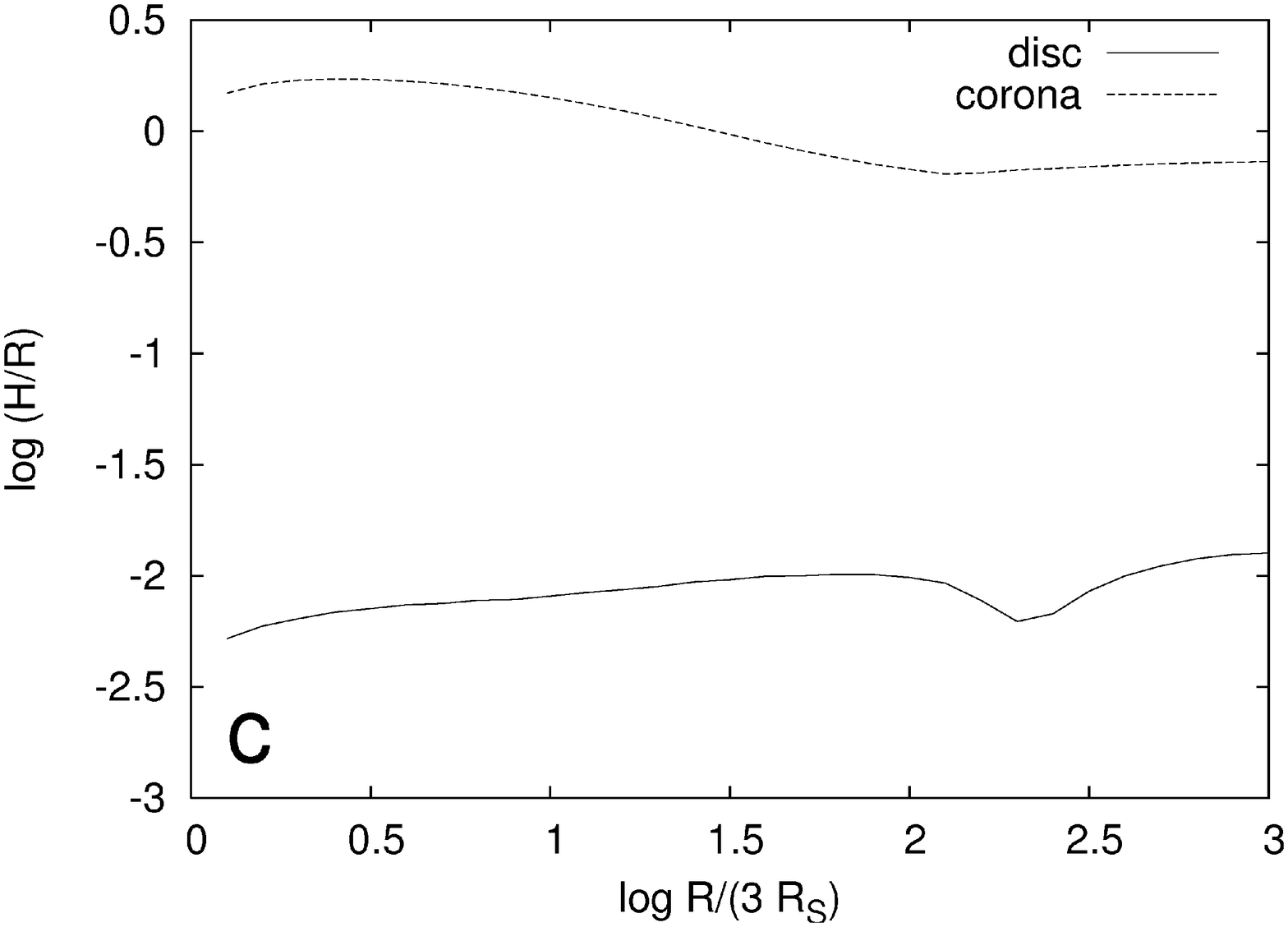} 
    &\includegraphics[width=0.48\textwidth]{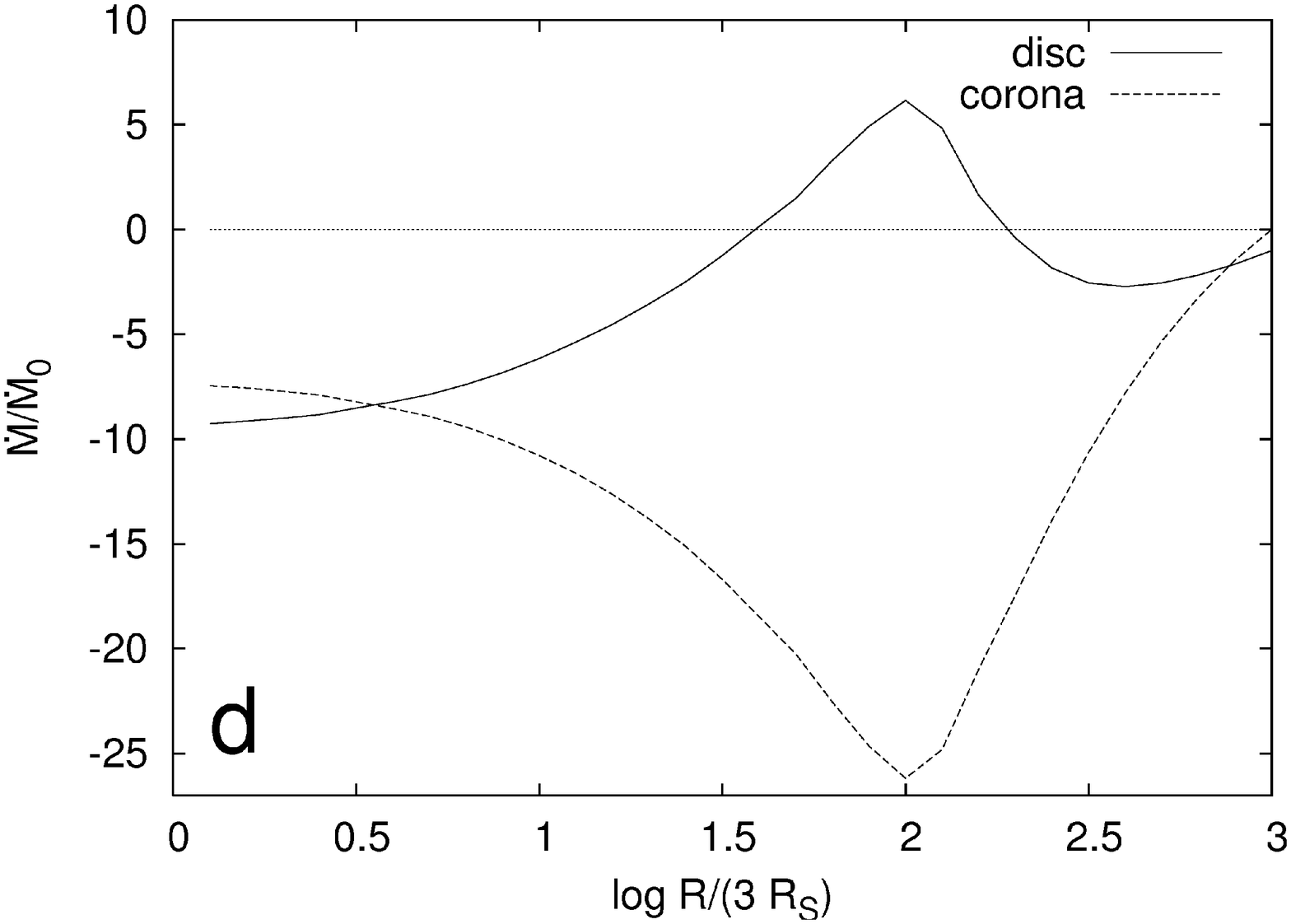}
  \end{tabular}
  \caption{The accretion flow 20 ks after the start: The corona
    production is at its maximum and so is the accretion rate in the
    corona. Most of the corona is produced at the transition from the
    two- to one-temperature plasma. As a result the disc flows
    outwards in some parts. Note that the ratio $H/R$ for the disc
    starts to decrease. The labelling is the same as in Fig.~\ref{fig:schemap1}.}
  \label{fig:schemap2}
\end{figure*}

\begin{figure*}
  \centering
  \begin{tabular}{cc}
    \includegraphics[width=0.48\textwidth]{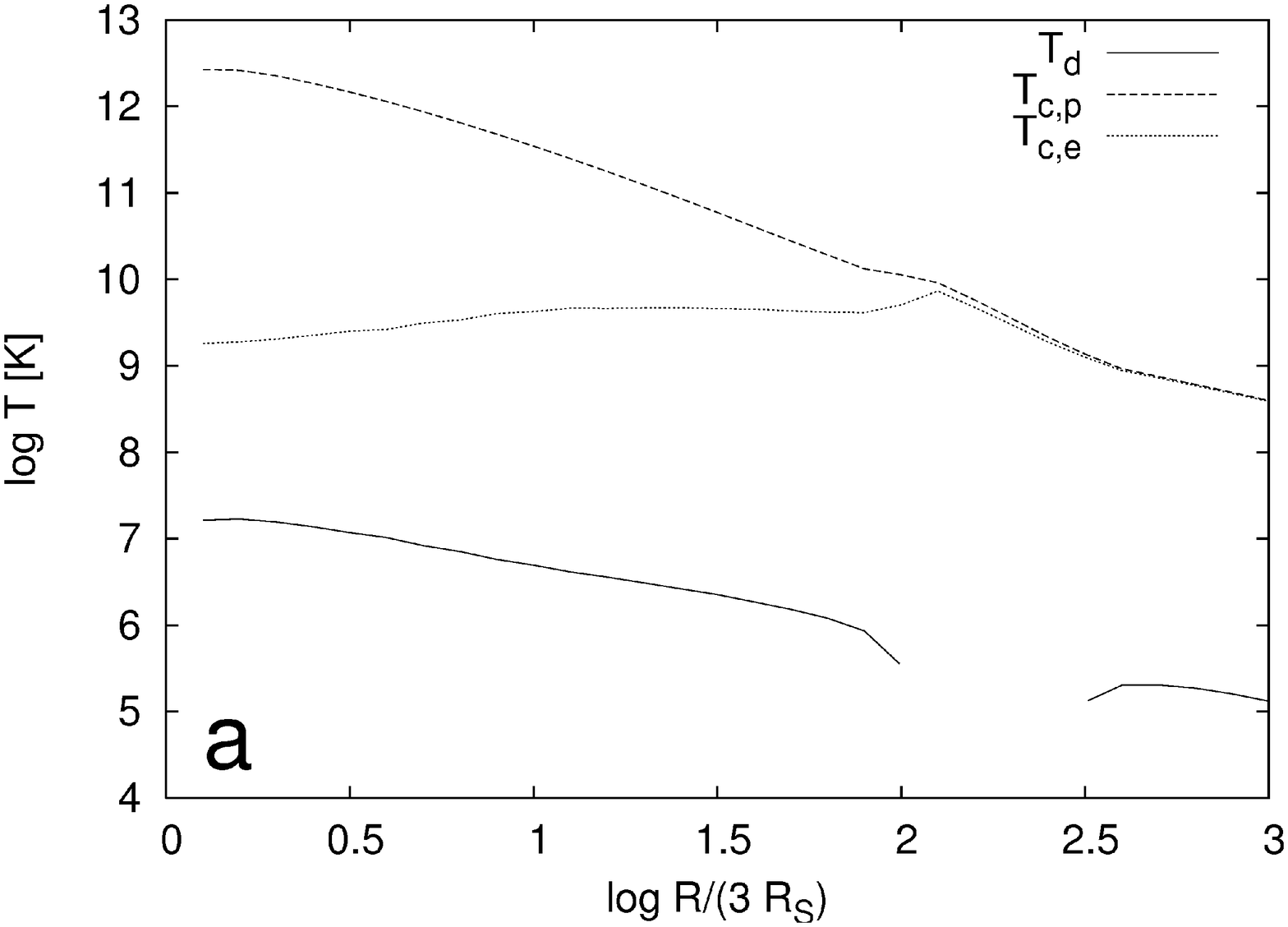}
    &\includegraphics[width=0.48\textwidth]{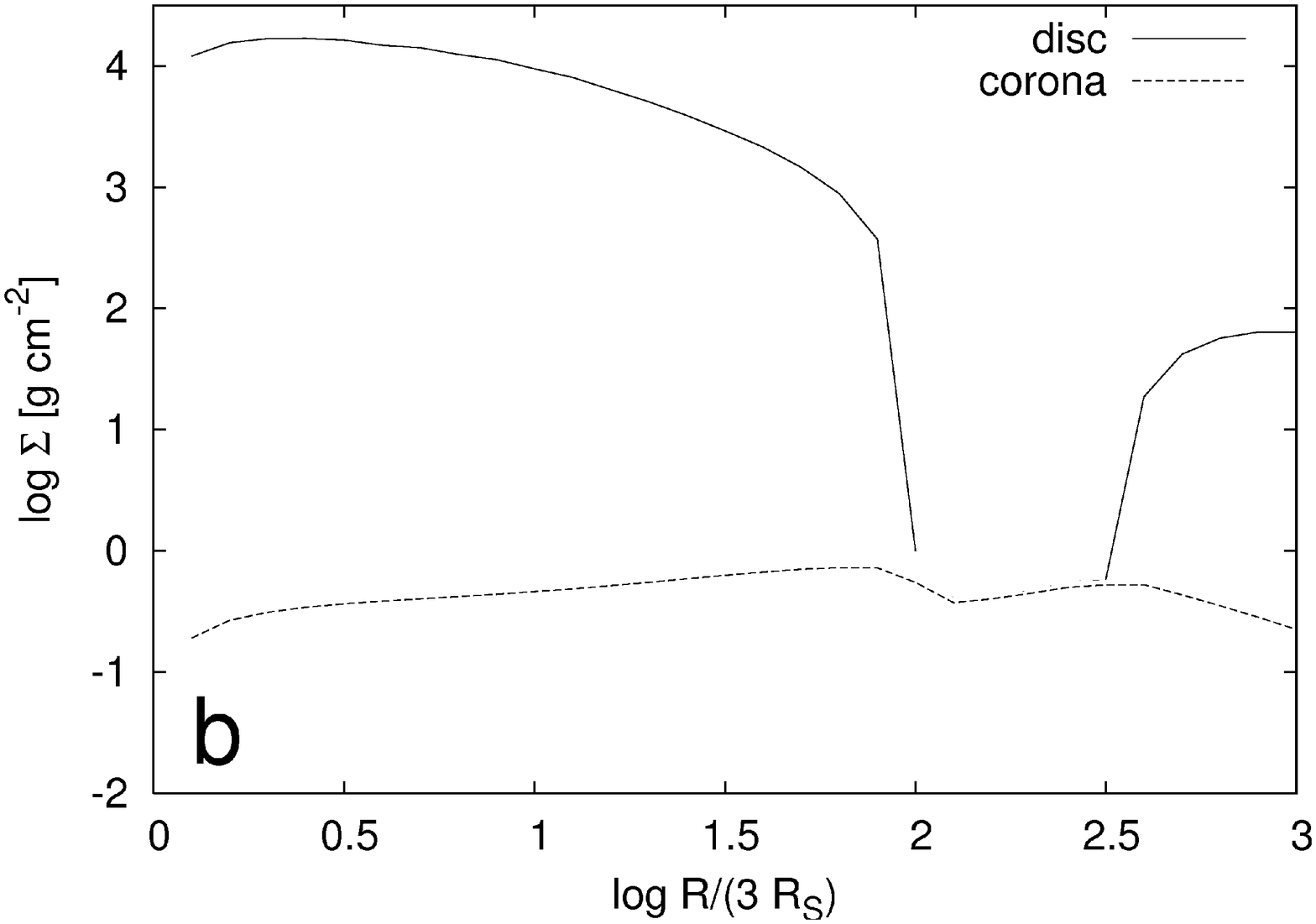}\\
    \includegraphics[width=0.48\textwidth]{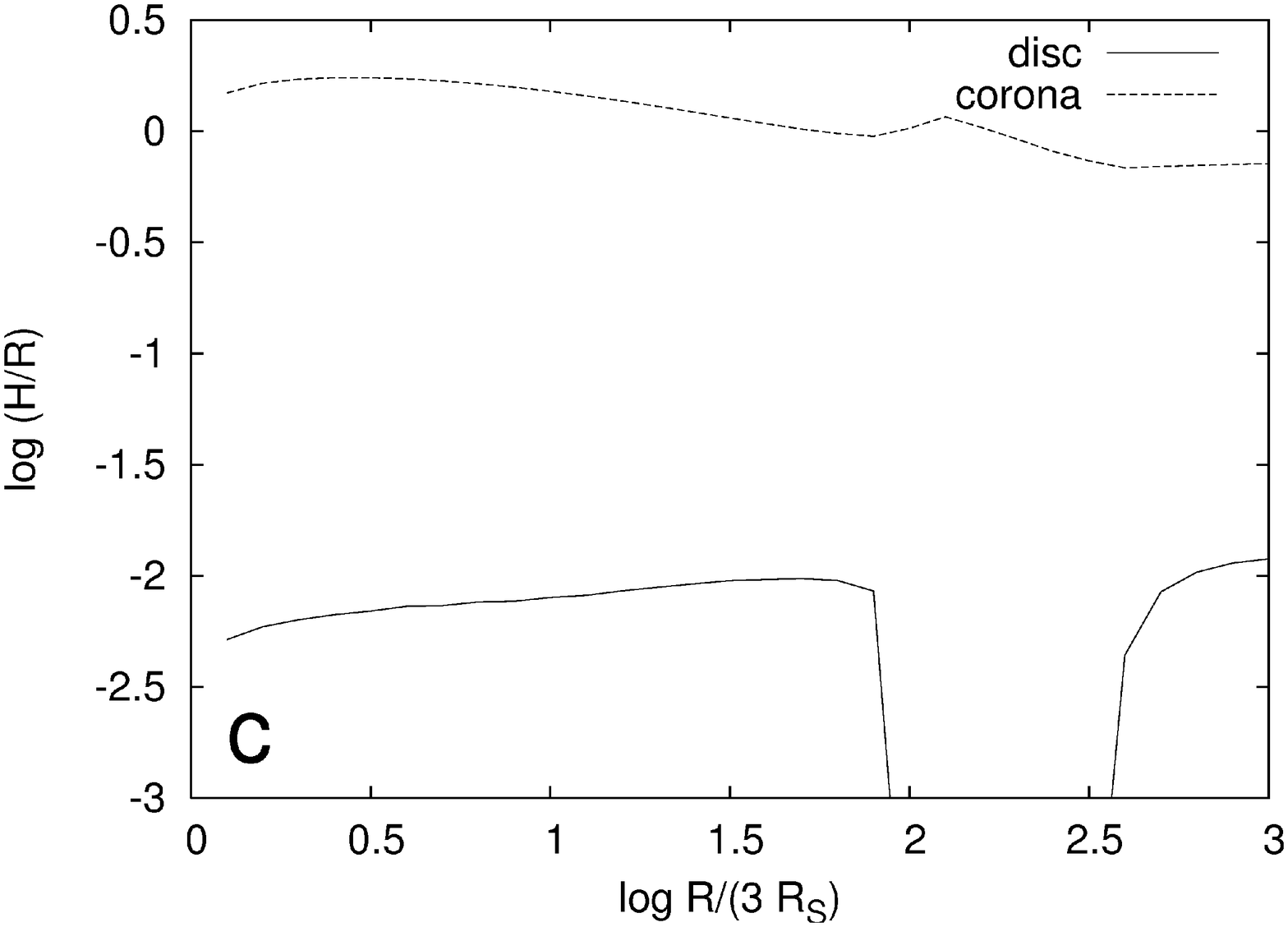} 
    &\includegraphics[width=0.48\textwidth]{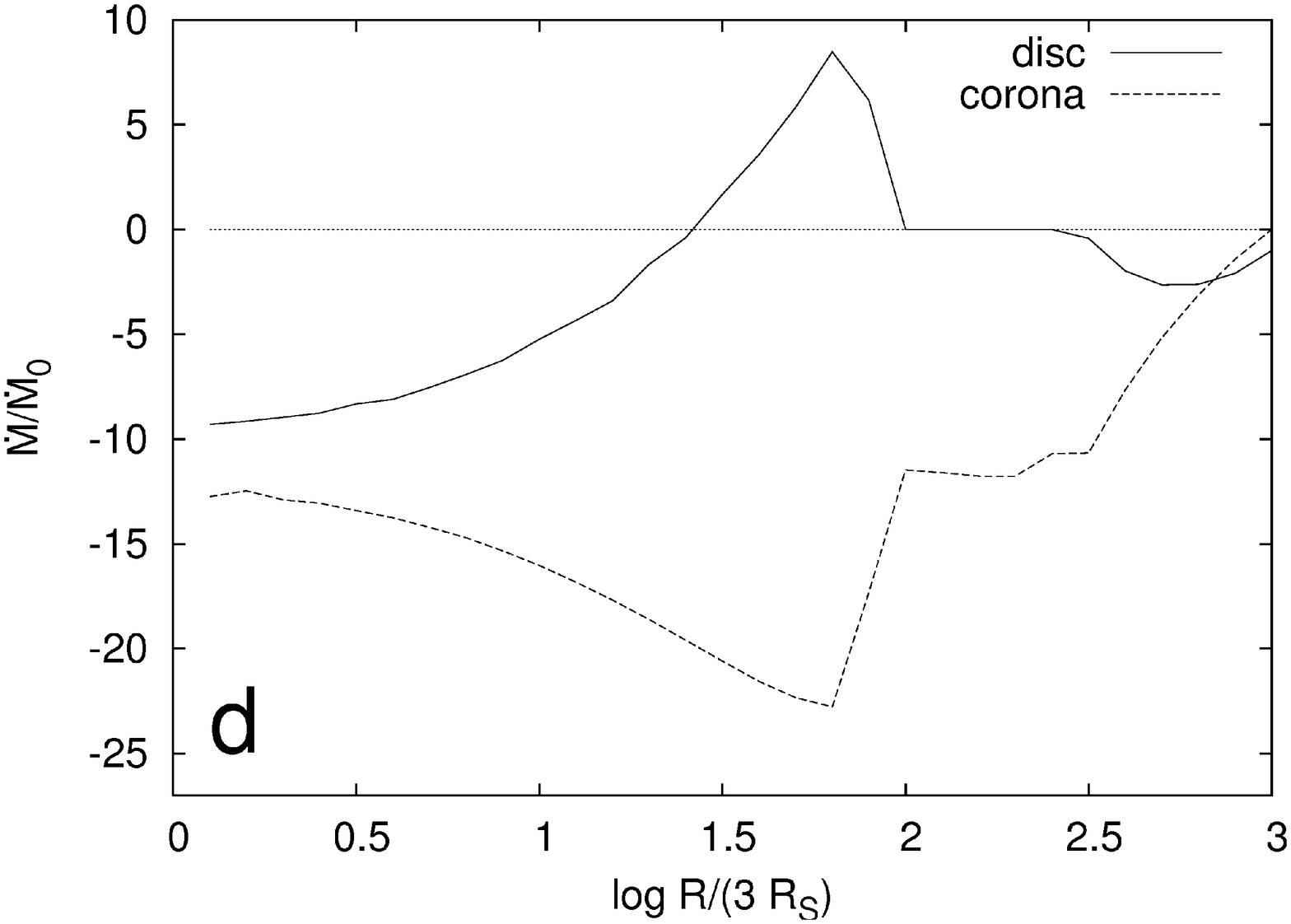}
  \end{tabular}
  \caption{The accretion flow 50 ks after the start: The disc develops
  a gap and the accretion flow in the disc is truncated. The inner
  disc disperses, i.e. parts flow outwards, while most of the matter
  is accreted. The outer disc evaporates and produces the
  corona. The coronal accretion flow in the gap is stationary,
  i.e. the accretion rate is radially constant (see Panel d) given its
much smaller viscous timescale. The labelling is the same as in Fig.~\ref{fig:schemap1}.}
  \label{fig:schemap3}
\end{figure*}

\begin{figure*}
  \centering
  \begin{tabular}{cc}
    \includegraphics[width=0.48\textwidth]{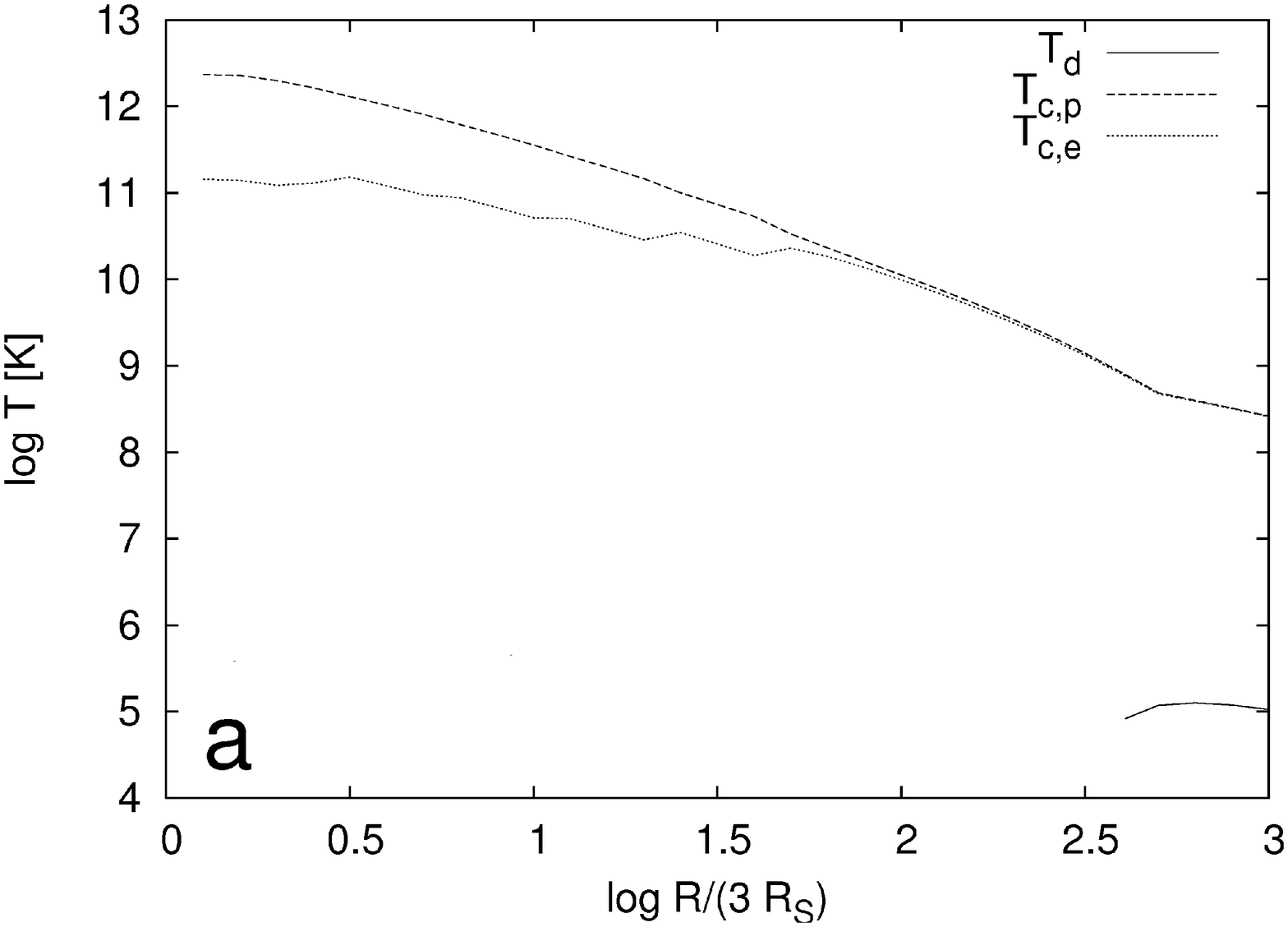}
    &\includegraphics[width=0.48\textwidth]{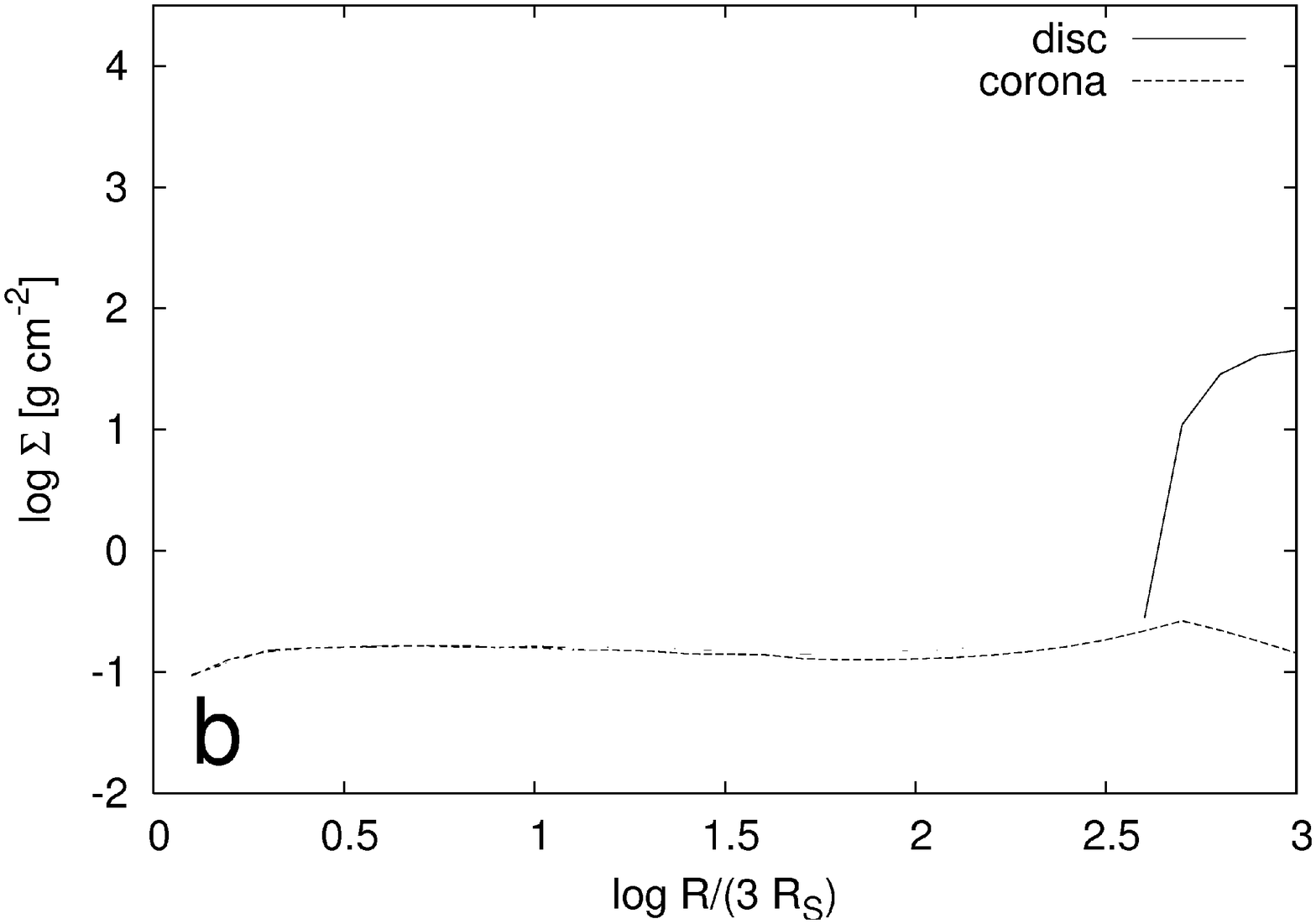}\\
    \includegraphics[width=0.48\textwidth]{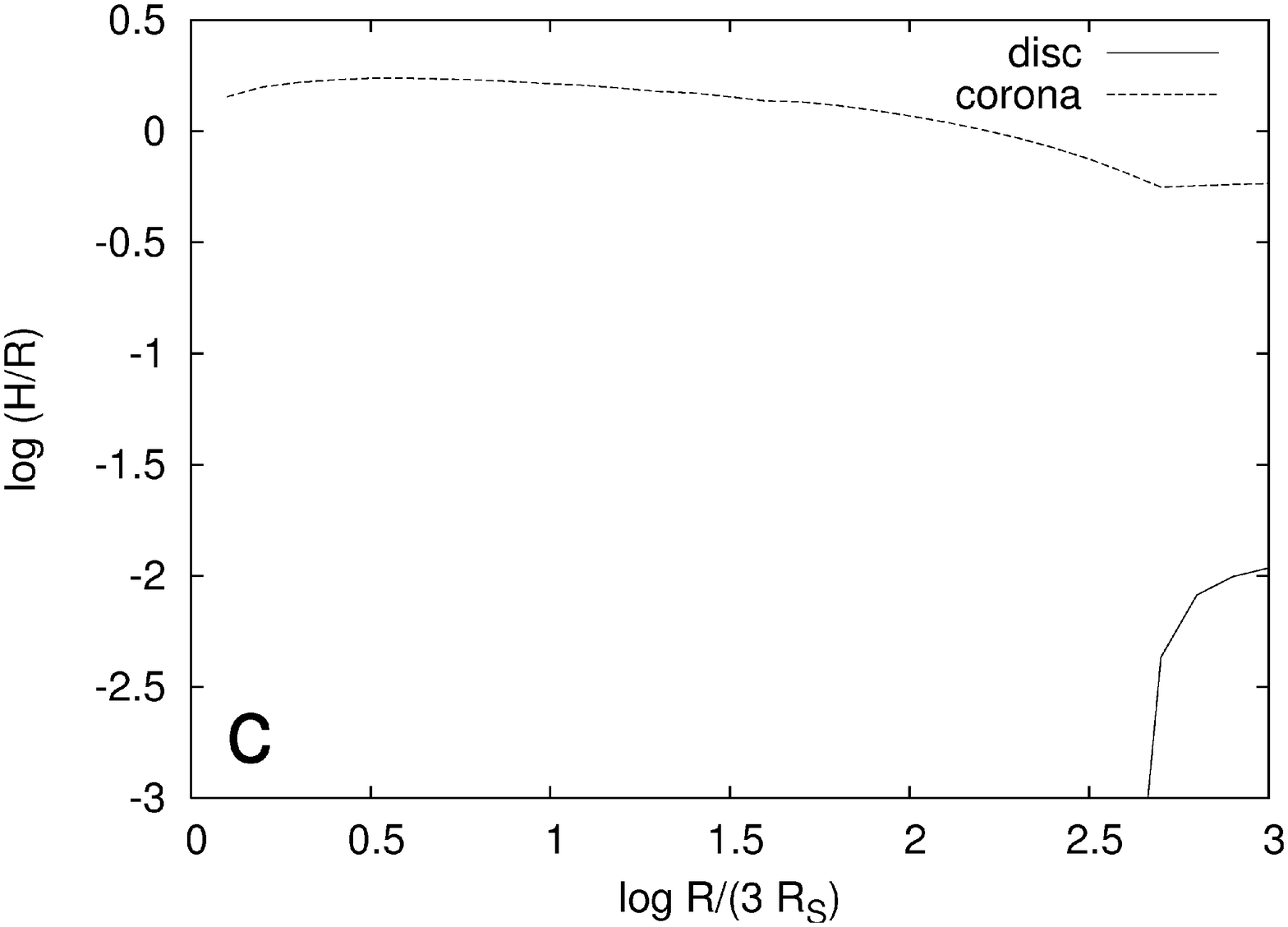} 
    &\includegraphics[width=0.48\textwidth]{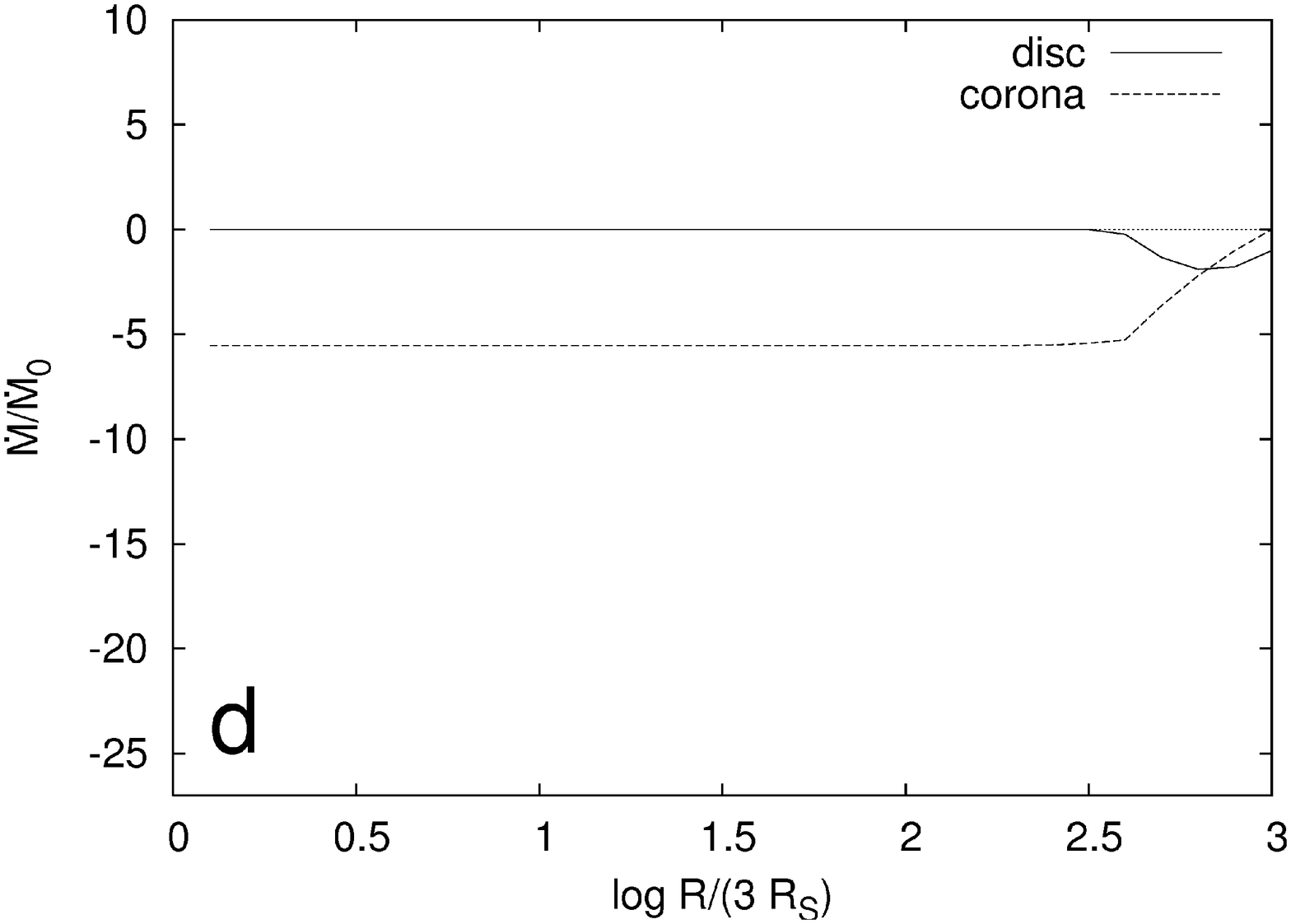}
  \end{tabular}
  \caption{The accretion flow 100 ks after the start: The inner disc
    completely dissolved by accretion and the inner parts of the flow
    consist only of a hot corona. This part of the corona is
    stationary. Further out there still is a remnant accretion disc
    producing the corona. The labelling is the same as in Fig.~\ref{fig:schemap1}.}
  \label{fig:schemap4}
\end{figure*}

In Fig.~\ref{fig:schemat} we show a schematic view of a typical
sandwich disc in the initial stage of the time-evolution. We show
relevant actual physical quantities 0.1, 20, 50 and 100
kiloseconds after the start of the simulation in 
Figs.~\ref{fig:schemap1}-\ref{fig:schemap4}. 

\begin{figure*}
  \centering
  
  \includegraphics[width=0.34\textwidth,angle=-90]{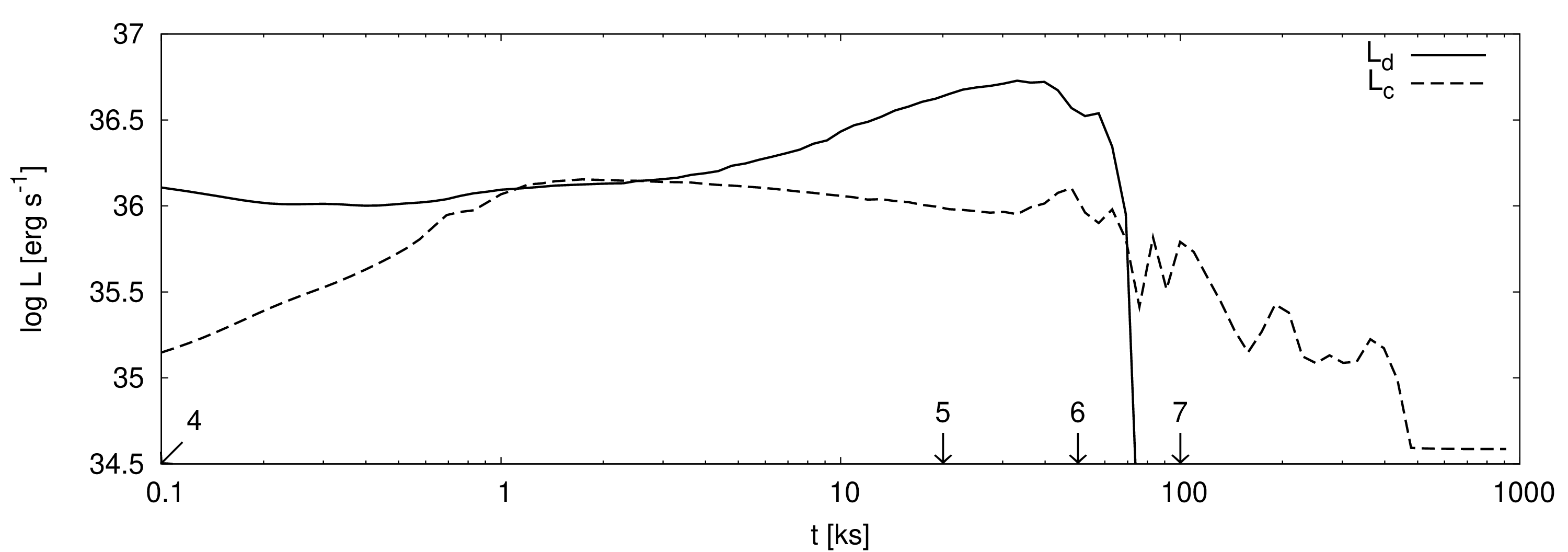}
  
  \caption{Luminosity of the disc $L_\textrm{d}$ and corona
    $L_\textrm{c}$. Initially the coronal luminosity increases quickly for
    1 ks. Then the corona condenses into the inner disc leading to an
    increase in the disc luminosity (cf. Fig.~\ref{fig:schemap1} and~\ref{fig:schemap2}). Starting at 40 ks, the disc
    luminosity declines as the gap widens. After about 70 ks the disc
    luminosity declines rapidly (The inner disc dissolves,
    cf. Fig.~\ref{fig:schemap3} and~\ref{fig:schemap4}) and the corona
    dominates the luminosity. The small-amplitude variations in the
    luminosity arise from the 
    coarse resolution and numerical fluctuations. The numbers annotated
    to the arrows on the time axis correspond to the figure number
    representing the respective state of the system. Note that we use a
    logarithmic scale for the time to differentiate the highlight the
    various timescales.}
  \label{fig:acclum}
\end{figure*}

We start with a stationary disc and corona extending over the total
computational domain. To compute the initial disc and corona structure
we neglect radial advection, and we take the fraction of accretion
occurring in the corona as $f_\textrm{c} = 0.1$.

Initially (Fig.~\ref{fig:schemap1}) in the outer regions the disc
begins to evaporate into the corona, whereas in the inner regions the
corona begins to condense into the disc (see Panel 1 in
Fig.~\ref{fig:schemat}). Since the disc loses a lot of matter in the
outer parts by evaporation into the corona, the corona temporarily
starts to flow outwards (Fig.~\ref{fig:schemap1}d). Most of the matter
is in the disc, and the corona is two-temperature inside a radius of
$R \approx 100~R_\textrm{S}$. 

Although the corona close to the black hole has got a very small
surface density in the initial model, as the outer disc however
evaporates mass into the corona, this matter is then transported
inwards through the corona. Further in, the coronal accretion rate
decreases as part of the coronal matter condenses into the disc and
fills up the corona to surface densities of about $0.3\dots 1$ g
cm$^{-2}$. This process happens very quickly, i.e. on the viscous
timescale of the corona. The increase in coronal mass in the inner
regions leads to an increase in the coronal luminosity. The inner
corona itself condenses into the disc and increases the disc surface
density (compare Fig.~\ref{fig:schemap1}b
and~\ref{fig:schemap2}b). Since most of the matter condenses very
close to the actual transition radius between the evaporating disc and
recondensing corona (which starts at around $R \approx
300~R_\textrm{S}$ and increases eventually to around $R \approx
1000~R_\textrm{S}$), it takes about a viscous timescale of the disc at
that radius to increase the surface density of the inner disc. This
timescale is $10^4$ times longer compared to the viscous timescale of
the corona.

At the transition radius (where the flow changes from an evaporating
disc to and recondensing corona), the disc starts to flow outwards
(see Panel 1 in Fig.~\ref{fig:schemat} and Fig.~\ref{fig:schemap2}d).
The accretion flow in the disc is then divided into three zones. In
the inner zone the disc is moving inwards, in the intermediate zone it
is moving outwards, and all the accretion is carried by the corona. In
the outer zone the disc is moving inwards. At the boundary of the
intermediate and the outer zone, despite the fact that the disc flow
is towards this radius, the surface density of the disc declines
rapidly. This comes about because most of the mass flowing into this
region from both sides is quickly evaporated into the corona.  An
empty ring, or gap, develops in the disc at this point. This gap
consists of coronal gas only and is Compton cooled by soft photons
from both the outer and inner disc remnants.

Once this gap has formed (Fig.~\ref{fig:schemap3}), the inner disc is
accreted and the gap widens. A decline in the disc scale-height is
already visible in Fig.~\ref{fig:schemap2}c, but the truncation of the
disc is fully evident in Fig.~\ref{fig:schemap3}. The inner disc
disperses. This state is shown schematically in Panel 2 of
Fig.~\ref{fig:schemat}.

Most of the inner disc gets accreted while a small part flows outwards
(see Fig.~\ref{fig:schemap3}d). The luminosity of the disc
declines. Eventually we are left with an accretion flow where the
inner part consists of a corona only and the mass supply for this
corona is produced by the outer remnant disc (cf. Panel 3 of
Fig.~\ref{fig:schemat} and Fig.~\ref{fig:schemap4}).

We show the luminosity evolution of the corona and disc in
Fig.~\ref{fig:acclum}. Initially, the luminosity of the corona
increases very quickly (caused as we described above by a filling of
the inner parts of the corona on the short viscous timescale of the
corona at the transition from the two- to a one-temperature
plasma), but cannot exceed the disc luminosity as long as a disc
exists (see Sect.~\ref{sect:importance}). 
On the much longer, viscous timescale of the disc the disc
luminosity increases (causes by a steady increase of surface density
of the inner disc), before the luminosity drops with the development
of the gap. Once the inner disc is completely accreted, the dominant
part of the radiation comes from the hot corona. The disc becomes
steady-state at this point, and no significant further evolution occurs.

\section{Discussion} \label{sect:discussion}

The results of these computations indicate that a disc around a 10
$M_\odot$ black hole, accreting at $\dot M = 0.001~M_\textrm{Edd}$
relaxes to a state in which the outer disc is a standard accretion
disc with a small corona, but the inner regions ($R \lesssim
1000~R_\textrm{S}$) consist of only hot coronal gas which has
evaporated from the disc.

\subsection{Thermal conduction and the transition from condensation to
evaporation} \label{sect:evapcrit}

These calculations basically agree with the results of
\citet{2000A&A...360.1170R} who concluded that for a stationary disc
there would be net condensation of the corona in the inner disc and
net evaporation further out. This comes about for the following
reasons.

The calculation of the evaporation/condensation rate
(eq.~\ref{eq:evapcond}) 
shows that the sign of the term $(q_\textrm{TL}^+-q_\textrm{TL}^-)$
determines condensation and evaporation, while the
$\kappa_\textrm{cond}$ term given its strong temperature dependence
determines the magnitude of the condensation/evaporation. This strong
temperature dependence also ensures that for the heating and cooling
mechanisms considered in this paper the mass transfer rate mainly
depends on the upper boundary of the integration, i.e. the coronal
values for the temperature. 

To get an idea what causes the evaporation/condensation in the two
zones shown in Fig.~\ref{fig:schemat} we assume that the only heating
is the $\alpha P_\textrm{c}$ heating (\ref{eq:qplusc}) and Compton cooling
(\ref{eq:comptcoolctl}) and bremsstrahlung are the only cooling mechanisms.

Within the corona the system is usually in thermodynamical equilibrium
or quickly relaxes back to equilibriumm -- faster closer to the black hole
than further out.  

In the two-temperature plasma (zone I) Compton cooling dominates. If
we calculate the difference between volume heating and cooling, we get
\begin{equation}
  \label{eq:heatcool1}
  q^+_\textrm{c}-q^-_\textrm{c,C}=P_\textrm{c}\left(\alpha_\textrm{c}\Omega_\textrm{c}-8\kappa_\textrm{es}\mu m_p\frac{T_\textrm{e}}{T_\textrm{e}+T_\textrm{p}}\frac{\sigma T_\textrm{eff,d}^4}{m_\textrm{e}c^2}\right)\;,
\end{equation}
where we have neglected Compton heating and the relativistic
generalisation in (\ref{eq:comptcoolctl}). Within the corona, this term is
about zero (assuming thermal equilibrium). To get the
evaporation/condensation rate, we need to integrate the difference
over the range of temperatures from the disc to the coronal values. If
the corona is a two-temperature plasma, then the fraction
$T_\textrm{e}/(T_\textrm{e}+T_\textrm{p})$ is small, but the closer we
get to the one-temperature plasma within the transition layer, the
bigger it becomes. Then the difference~(\ref{eq:heatcool1}) becomes
increasingly negative. The corona condensates onto the disc. 

In zone II we have got a one-temperature plasma and Bremsstrahlung dominates the cooling. We need to calculate
(For simplicity we use the classical Bremsstrahlung result,
i.e. $q_\textrm{c,brems}\propto \rho^2T^{1/2}$)
\begin{equation}
  \label{eq:heatcool2}
  q^+_\textrm{c}-q^-_\textrm{c,brems}=P_\textrm{c}\left(\alpha_\textrm{c}\Omega_\textrm{c}-C\frac{P_\textrm{c}}{T^{3/2}}\right)\;,
\end{equation}
where $C$ is an arbitrary constant of proportionality.
While the difference is zero in the corona, the further down we
integrate towards the disc temperature, the smaller gets $T$ and the
larger the second term in the difference. Cooling dominates heating in
the transition layer and thus the corona condenses into the disc. This
is in contrast with our results presented in
Section~\ref{sect:results}. The outer parts of the disc evaporate
because the outward advection in the initial phases of the disc
evolution lead to a imbalance between heating and cooling. Heating
dominates there and hence the disc evaporates.

\subsection{Physical processes neglected in this model}

We now discuss some of the additional physical processes which we have
neglected here, but which will need to be taken into account if the
model is to be developed further.

\subsubsection{Other cooling mechanisms} \label{sect:othercool}

Further out other cooling mechanisms than Compton cooling and
bremsstrahlung will be more efficient and eventually lead to a
condensation of the corona. We neglect atomic cooling and only allow
for bremsstrahlung and Compton cooling for both the corona and the
transition layer. The temperature dependence of bremsstrahlung and
Compton cooling ensures that the evaporation/condensation rate
(cf. eq.~\ref{eq:evapcond}) only depends on the upper limits of the
integral involved. If we needed to include cooling processes other
than bremsstrahlung (i.e. atomic cooling etc.), then the calculation
of the evaporation/condensation rate would be much more difficult. In
the transition layer the gas then can have a significant absorption
optical depth which in turn leads to photoionisation processes and
significant radiation pressure. Photoionisation and radiation pressure
also influence the corona if it starts to collapse and may lead
to an outflow (See Sect.~\ref{sect:windflow}). The increase in cooling
may also be responsible for changing the sign of $\dot m_z$ from
evaporation to condensation in the outer parts of the disc and thus
will stop the transition radius to move outwards.

With the widening of the inner gap a lack of soft photons develops and
the electron temperature reaches temperatures in excess of $10^{10}$
K. This invalidates our use of non-relativistic Bremsstrahlung
rates. However given the other underlying approximations in the model,
the missing factors of a few should not influence the results too
much. In a more realistic model, pair production will set in
\citep[e.g.][]{1982ApJ...258..335S} and help to cool the corona. Pair
production increases the electron scattering optical depth and thus
also leads to a rise in the Compton cooling rate. In fact, many, if
not all observations of AGN hint at the existence of a population of
hot electrons in a very narrow range of electron temperatures (about
100 keV, which corresponds to $10^9$ K \citep[for NGC 4151,
see][]{2005ApJ...634..939B}. These authors deduce an optical depth for
the corona of $\tau\approx 1.3$. This is much higher than the values
encountered in the model presented in this paper.

The study of these cases is very important but requires further work.

\subsubsection{Wind/Outflow} \label{sect:windflow}

We have neglected the possibility of a wind/outflow throughout these
calculations. Physically given the large ratio of $H/R$ for the
corona, the corona is only marginally bound within the gravitational
potential \citep{1999MNRAS.303L...1B}. Thus it may well be that mass,
energy and angular momentum is carried away in an outflow, jet or
wind. This might then eventually lead to a reduction of $H/R$ until a
stable state is found. This would then be an additional cooling
mechanism for the corona. Observations of low-luminosity AGN and X-Ray
binaries in their low/hard state indicate that they are dominated by
large-scale outflows \citep{2004MNRAS.355.1105F}. These outflows
can be driven magnetically \citep[e.g.][]{2006MNRAS.368..379M} or by
radiation pressure as well (cf. Sect.~\ref{sect:othercool}).

\subsubsection{Corona production} \label{sect:prodcor}

Another important question to be answered is the production of the
corona. Our model presented in this paper shows that once we start
with a tiny corona and consider thermal conduction, the corona is
produced mainly from the outer optically thick disc. How the corona
came into existence in the first place is not clear. As described in
Sect.~\ref{sect:trunc}, we have run simulations for a one-phase black
hole accretion disc  but allowing for the transition to an
optically thin flow. The simulations are similar to the ones reported
in \citet{2006MNRAS.368..379M}, except that we now use a prescription
by \citet{1996ApJ...456..119A} for the optically thick-thin
transition. As reported before, we still find limit cycles for the
same range of accretion rates. During the high-$\dot M$ phase, where
the inner accretion disc is depleted, this part of the accretion flow
becomes optically thin and the temperature rises to about $10^9$ K
close to the black hole. This might be a promising mechanism to
produce a corona and a possible explanation for the so-called very high
(VH) or steep power law (SPL) state in X-Ray binaries
\citep{2006ARA&A..44...49R}.  This optically thin coronal part (being
geometrically thick) then can flow out-/inwards on top of the
accretion disc.

\subsubsection{Extra shear}

In the course of this paper, we neglect non-Keplerian rotation for the
corona. We here discuss some possible effects of the inclusion of
non-Keplerian rotation into the model. These could play an important
role in determining energy dissipation and the redistribution of
angular momentum (i.e. accretion) in the disc and the corona. It will
be important to investigate the outcome of including these effects in
future computations.

Since the corona is geometrically thick, radial pressure gradients
lead to a deviation from pure Keplerian rotation. Corona and disc thus
rotate at different speeds. At their boundary (the transition layer)
torques arise which lead to angular momentum exchange between the two
phases. The work done by these torques leads to an additional heating
of the corona, in particular the ions. Further angular momentum
exchange arises from the mass transfer due to thermal conduction as
then both layer rotate on different speeds. We give here a rough
calculation of the torques and dissipation rates which might arise
from these processes and the effect of the extra torque on
the accretion flow.

The rotation frequency $\Omega_\textrm{c}\neq
\Omega_\textrm{K}=\sqrt{GM/R^3}$ in the presence of a radial pressure
gradient is given by
\begin{equation}
  \label{eq:omegapress}
  \Omega_\textrm{c}=\sqrt{\Omega_\textrm{K}^2+\frac{1}{\rho_\textrm{c} R}\frac{\partial
      P_\textrm{c}}{\partial R}}\;,
\end{equation}
where $\rho_\textrm{c}$ is the coronal density, $P_\textrm{c}$ the
pressure and $R$ the radial distance from the black hole.  We
approximate the pressure gradient by $\partial P_\textrm{c}/\partial
R=P/R\chi_P$ with $\chi_P=\partial \log P_\textrm{c}/\partial \log
R$. Since $P_\textrm{c}\approx \rho_\textrm{c} c_\textrm{s,c}^2$ and
$c_\textrm{s,c}\approx \Omega_\textrm{c} H_\textrm{c}$, where
$c_\textrm{s,c}$ is the sound speed and $H_\textrm{c}$ is the
scale-height of the corona, we get
\begin{equation}
  \label{eq:omegapress2}
  \Omega_\textrm{c}=\frac{\Omega_\textrm{K}}{\sqrt{1-\left(H_\textrm{c}/R\right)^2\chi_P}}\;.
\end{equation}
If $\Delta\Omega=\Omega_\textrm{K}-\Omega_\textrm{c}$, then for small $H_\textrm{c}/R<1$ we
can write
\begin{equation}
  \label{eq:deltaomega}
  \frac{\Delta \Omega}{\Omega_\textrm{K}}=-\frac{1}{2}\left(\frac{H_\textrm{c}}{R}\right)^2\chi_P\;.
\end{equation}
Thus coronal and disc flow rotate at different speed. The differential
rotation leads to a free energy per unit surface area of about
\begin{equation}
  \label{eq:free-e}
  \Delta
  E_\textrm{c}=\frac{1}{2}\frac{\Sigma_\textrm{d}\Sigma_\textrm{c}}{\Sigma_\textrm{d}+\Sigma_\textrm{c}}\left(\Delta \Omega R\right)^2\;. 
\end{equation}
We note that usually $\Sigma_\textrm{d}\gg \Sigma_\textrm{c}$ and thus $\Delta
  E_\textrm{c}\approx \frac{1}{2}\Sigma_\textrm{c}\left(\Delta \Omega
    R\right)^2$. This energy is released on a multiple of the local dynamical
timescale, i.e. we set
\begin{equation}
  \label{eq:t-e}
  \Delta t_{E_\textrm{c}}=f_{E_\textrm{c}}\Omega_\textrm{c}^{-1}\;,
\end{equation}
where $f_{E_\textrm{c}}\geq 1$ is a parameter and
$\Omega_\textrm{c}^{-1}$ the dynamical timescale. The release of the
free energy (\ref{eq:free-e}) leads to an additional heating of the
corona at a rate
\begin{equation}
  \label{eq:heatcordiff}
  Q_\textrm{c,diff}\equiv \frac{\Delta E}{\Delta
    t}=\frac{1}{2}\frac{\Sigma_\textrm{d}\Sigma_\textrm{c}}{\Sigma_\textrm{d}+\Sigma_\textrm{c}}\left(\Delta \Omega R\right)^2\Omega_\textrm{c}f_{E_\textrm{c}}^{-1}\;.
\end{equation}
If we put~(\ref{eq:deltaomega}) into the last equation, then we get
\begin{equation}
  \label{eq:heatcordiff2}
  Q_\textrm{c,diff}=\frac{1}{8}\frac{\Sigma_\textrm{d}\Sigma_\textrm{c}}{\Sigma_\textrm{d}+\Sigma_\textrm{c}}\left(\Omega_\textrm{K} R\right)^2 \left(\frac{H_\textrm{c}}{R}\right)^4\chi_P^2\Omega_\textrm{c}f_{E_\textrm{c}}^{-1}\;.
\end{equation}
The heating timescale for this process is
\begin{equation}
  \label{eq:tscaleheat-diff}
  \tau_\textrm{diff,c}\equiv\frac{\Sigma_\textrm{c}
    c_\textrm{s,c}^2}{Q_\textrm{c,diff}}=8\frac{\Sigma_\textrm{d}+\Sigma_\textrm{c}}{\Sigma_\textrm{d}}\left(\frac{R}{H_\textrm{c}}\right)^2\left(\frac{\Omega_\textrm{c}}{\Omega_\textrm{K}}\right)^2 \chi_P^{-2}f_{E_\textrm{c}}\Omega_\textrm{c}^{-1}\;. 
\end{equation}
Thus the heating timescale is longer than the thermal timescale
$\tau_\textrm{c}$ of the corona ($\tau_\textrm{c}\approx
(\alpha_\textrm{c}\Omega_\textrm{c})^{-1}$) for  
\begin{equation}
  \label{eq:diffcond}
  f_{E_\textrm{c}}>\frac{1}{8\alpha_\textrm{c}}\left(\frac{\Omega_\textrm{K}}{\Omega_\textrm{c}}\right)^2\left(\frac{H_\textrm{c}}{R}\right)^2\chi_P^{2}\;.
\end{equation}

The differential rotation leads to a torque which tries to speed up
the corona and slow down the disc, i.e. the torque is
\begin{equation}
  \label{eq:torquediff}
  G_\textrm{diff}\approx Q_\textrm{c,diff}/\Omega_\textrm{c}=\frac{1}{8}\frac{\Sigma_\textrm{d}\Sigma_\textrm{c}}{\Sigma_\textrm{d}+\Sigma_\textrm{c}}\left(\Omega_\textrm{K} R\right)^2 \left(\frac{H_\textrm{c}}{R}\right)^4\chi_P^{-2}f_{E_\textrm{c}}^{-1}\;.
\end{equation}
If only the torques given in eqns.~(\ref{eq:torquecorona})
and~(\ref{eq:torquedisc}) would be at work in the corona and disc,
respectively, then in a stationary Keplerian disc (neglecting mass exchange due
to thermal conduction) matter flows at a speed 
\begin{equation}
  \label{eq:vrdisc}
  u_{R,\textrm{visc}}=\frac{1}{2}\alpha c_\textrm{s}\left(\frac{H}{R}\right)\chi_\Omega\approx \frac{1}{2}\Omega R\left(\frac{H}{R}\right)^2\chi_\Omega\;.
\end{equation}
As long as $\Omega$ decreases with increasing radial distance,
i.e. $\chi_\Omega=(\partial\log \Omega)/(\partial \log R)\approx -3/2<0$, then
matter always flows inwards as expected. If we include the effects of the extra
torque~(\ref{eq:torquediff}), then the contribution to the speed of
matter is
\begin{equation}
  \label{eq:vrdiff}
  u_{R,\textrm{diff}}=\frac{G_\textrm{diff}}{\Sigma R
    \Omega\left(2+\chi_\Omega\right)}\;.
\end{equation}
For a near Keplerian rotating flow we approximate $\chi_\Omega= -3/2$
and put in eq.~(\ref{eq:torquediff}) to get
\begin{equation}
  \label{eq:vrdiffc}
  u_{R,\textrm{diff,c}}=+\frac{1}{8}\frac{\Sigma_\textrm{d}}{\Sigma_\textrm{d}+\Sigma_\textrm{c}}\left(\frac{\Omega_\textrm{K}}{\Omega_\textrm{c}}\right)^2 \left(\frac{H_\textrm{c}}{R}\right)^4\chi_P^2f_{E_\textrm{c}}^{-1} \Omega_\textrm{c} R
\end{equation}
and
\begin{equation}
  \label{eq:vrdiffd}
  u_{R,\textrm{diff,d}}=-\frac{1}{8}\frac{\Sigma_\textrm{c}}{\Sigma_\textrm{d}+\Sigma_\textrm{c}} \left(\frac{H_\textrm{c}}{R}\right)^4\chi_P^2f_{E_\textrm{c}}^{-1} \Omega_\textrm{K} R
\end{equation}
for corona and disc, respectively. The relative contribution to the
inflow caused by the viscous torques are
\begin{equation}
  \label{eq:vrdiffc2}
\frac{u_{R,\textrm{diff,c}}}{u_{R,\textrm{visc,c}}}=-\frac{1}{4\alpha_\textrm{c}}\frac{\Sigma_\textrm{d}}{\Sigma_\textrm{d}+\Sigma_\textrm{c}}\left(\frac{\Omega_\textrm{K}}{\Omega_\textrm{c}}\right) \left(\frac{H_\textrm{c}}{R}\right)^2\chi_P^2f_{E_\textrm{c}}^{-1} 
\end{equation}
and
\begin{equation}
  \label{eq:vrdiffd2}
\frac{u_{R,\textrm{diff,d}}}{u_{R,\textrm{visc,d}}}=\frac{1}{4\alpha_\textrm{d}}\frac{\Sigma_\textrm{c}}{\Sigma_\textrm{d}+\Sigma_\textrm{c}}\left(\frac{\Omega_\textrm{K}}{\Omega_\textrm{c}}\right) \left(\frac{H_\textrm{c}}{R}\right)^2\left(\frac{H_\textrm{c}}{H_\textrm{d}}\right)^2\chi_P^2f_{E_\textrm{c}}^{-1} \;.
\end{equation}
Thus for 
\begin{equation}
  \label{eq:fecest}
  f_{E_\textrm{c}}<\frac{1}{4\alpha_\textrm{c}}\frac{\Sigma_\textrm{d}}{\Sigma_\textrm{d}+\Sigma_\textrm{c}}\left(\frac{\Omega_\textrm{K}}{\Omega_\textrm{c}}\right) \left(\frac{H_\textrm{c}}{R}\right)^2\chi_P^2
\end{equation}
the corona starts to flow outwards and for 
\begin{equation}
  \label{eq:fedest}
  f_{E_\textrm{c}}<\frac{1}{4\alpha_\textrm{d}}\frac{\Sigma_\textrm{c}}{\Sigma_\textrm{d}+\Sigma_\textrm{c}}\left(\frac{\Omega_\textrm{K}}{\Omega_\textrm{c}}\right) \left(\frac{H_\textrm{c}}{R}\right)^2\left(\frac{H_\textrm{c}}{H_\textrm{d}}\right)^2\chi_P^2
\end{equation}
the inflow speed of the disc is dominated by the extra torque. 

For a typical disc and corona, we have $\chi_P\approx -1$,
$H_\textrm{c}/R\approx 1$, $H_\textrm{d}/R\approx 10^{-2}$,
$\Omega_\textrm{c}\approx \Omega_K/\sqrt{2}$ and $\Sigma_\textrm{d}\gg
\Sigma_\textrm{c}$, then we have
\begin{itemize}
\item an outflowing corona for
$f_{E_\textrm{c}}<1/(4\alpha_\textrm{c}\sqrt{2})$ 
\item and an inflowing disc
dominated by the extra torque for
$f_{E_\textrm{c}}<10^4/(4\alpha_\textrm{d}\sqrt{2})$.
\item Since we physically cannot resolve timescales smaller than the dynamical
timescale, i.e. $f_{E_\textrm{c}}\ge 1$, $\alpha_\textrm{c}$ has to be smaller
than $\alpha_\textrm{c}<1/(4\sqrt{2})\approx 0.18$ for both effects to
work. 
\end{itemize}

If we include the effects of thermal conduction, then the limit on
$f_{E_\textrm{c}}$ for the outflow of the corona decreases(increases)
if the disc evaporates(corona condenses), since then the mass exchange
transfers angular momentum as well. If the disc evaporates ($\dot
m_z>0$), the corona yields angular momentum at a rate of $\dot m_z
R^2\Delta \Omega$.

The extra torque can be produced by the poloidal component of the
magnetic field, $B_z$, in the disc which slightly overextends into the
corona. The smaller speed transforms the poloidal field into an
azimuthal field. It creates an twist leading to a azimuthal field of
the order of
\begin{equation}
  \label{eq:polfield}
  B_\phi=-B_z\frac{\Delta \Omega}{\Omega_\textrm{c}}
\end{equation}
after one orbit. Thus the torque produce by this process can be
expressed as
\begin{equation}
  \label{eq:tmag}
  T_\textrm{mag}=R\left(\frac{B_\phi B_z}{4\pi}\right)=R\frac{ B_z^2}{4\pi}\frac{\Delta \Omega}{\Omega_\textrm{c}}\;.
\end{equation}
We furthermore can assume that the poloidal magnetic field strength
$B_z$ is essentially equal to the intrinsic disc magnetic field
$B_\textrm{disc}$. Thus we slightly underestimate
$B_\textrm{disc}$. If we further use $B_\textrm{disc}^2=4\pi
\alpha_\textrm{d} P_\textrm{d}$ \citep{1973A&A....24..337S}, then we
can write for the torque in~(\ref{eq:tmag})
\begin{equation}
  \label{eq:tmag2}
  T_\textrm{mag}=R\frac{\Delta
    \Omega}{\Omega_\textrm{c}}\alpha_\textrm{d} P_\textrm{d}\;.
\end{equation}
For $\Omega_\textrm{c}=\Omega_\textrm{K}/\sqrt{2}$ we get with
eq.~(\ref{eq:deltaomega})
\begin{equation}
  \label{eq:tmag3}
  T_\textrm{mag}=-\sqrt{2}\left(\frac{H_\textrm{c}}{R}\right)^2\alpha_\textrm{d} P_\textrm{d}\chi_P\;.
\end{equation}
Finally we can use $P_\textrm{d}\approx \rho_\textrm{d}
c_\textrm{s,d}^2\approx \Sigma_\textrm{d}\Omega_\textrm{K}^2
H_\textrm{d}$ and arrive at
\begin{equation}
  \label{eq:tmag4}
  T_\textrm{mag}=-\alpha_\textrm{d}\sqrt{2}\left(\frac{H_\textrm{d}}{R}\right)\left(\frac{H_\textrm{c}}{R}\right)^2 \Sigma_\textrm{d} \left(\Omega_\textrm{K} R\right)^2 \chi_P\;.
\end{equation}
The comparison of~(\ref{eq:tmag3}) and~(\ref{eq:torquediff}) allows to
estimate $f_{E_\textrm{c}}$. We get
\begin{equation}
  \label{eq:fec}
  f_{E_\textrm{c}}=-\frac{1}{8\alpha_\textrm{d}\sqrt{2}}\frac{\Sigma_\textrm{c}}{\Sigma_\textrm{c}+\Sigma_\textrm{d}} \left(\frac{H_\textrm{c}}{R}\right)^2\chi_P^{-3} \left(\frac{H_\textrm{d}}{R}\right)^{-1}
\end{equation}
For typical values for the corona ($\chi_P=-1$, $H_\textrm{c}\approx
R$) and typical values for the disc ($\Sigma_\textrm{d}\approx
10^{2\dots 4} \Sigma_\textrm{c}$ and $H_\textrm{d}/R\approx 10^{-2}$)
we get $f_{E_\textrm{c}}\approx (10^{-2}\dots
1)/(8\sqrt{2}\alpha_\textrm{d})$. For $\alpha_\textrm{d}=0.1$ this
leads to $f_{E_\textrm{c}}\approx 10^{-2}\dots 1$. Since we
underestimate $B_\textrm{disc}$, another factor larger than unity
applies to the result in~(\ref{eq:fec}). Hence our estimate of
$f_{E_\textrm{c}}$ given the physical picture of the magnetic torque
discussed above is compatible with an outflowing corona. It however
constrains $f_{E_\textrm{c}}$ to be not significantly larger than
unity.

Note that the inclusion of thermal conduction introduces a
non-constant accretion rate $\dot M_\textrm{c}(R)$ and $\dot
M_\textrm{d}(R)$ for the corona and disc, respectively, even for the
stationary case.  Then extra
factors enter the estimations presented above.

The physical process presented here might be responsible for driving a
corona produced close to the black hole further outwards where it can
be influence the spectrum and may condense into the disc
(cf. Sect.~\ref{sect:prodcor}).

\section{Conclusions} \label{sect:conclusions}

We present time-dependent simulations for a two-phase accretion flow
around a black hole. We start with a flow which initially consists of
a cool and geometrically thin but optically thick accretion disc,
sandwiched by a hot and geometrically thick but optically thin corona.
We show that as the disc evolves, the disc-corona sandwich vanishes
inside a certain radius. A gap in the disc develops and the inner disc
is fully accreted. In the final state,  we are left with a coronal hot
inner flow and a disc-corona sandwich further out.

The general behaviour of the disc as described in this paper is not
completely unexpected, since there is mounting evidence that the
accretion flow around black holes at low accretion rates is truncated
inside some radius (see Sect.~\ref{sect:importance}). While we
assume a stellar mass black hole of 10 solar masses, the result
is likely also to apply to higher black hole masses, i.e. AGN. The
result is in support of the unified scenario of
\citet{1997ApJ...489..865E}, where the optically thick disc is
truncated below a certain radius for low accretion rates and only
reaches the last stable orbit for higher accretion rates close to the
Eddington limit.

For our model, we tried to keep the number of assumptions as small and
the physics as simple as possible. But clearly there are still areas
for improvement which we discuss in Section~\ref{sect:discussion} and
plan to include in the future work.  The model presented here
nevertheless gives an indication that the standard disc at low
accretion rates cannot exist close to the black hole.  The physical
reasons for this are discussed in Section~\ref{sect:evapcrit}.

\section*{Acknowledgements}

We gratefully acknowledge the use of a fast PC acquired for JEP by
Emmanuel College, Cambridge. MM acknowledges support from PPARC. MM
also thanks Massimo Cappi, Mauro Dadina and Gabriele Ponti for
interesting discussions.

\appendix
\onecolumn

\section{Transformations} \label{sect:trafo}

We define the specific energy per unit mass for the disc as
\begin{equation}
  \label{eq:ed}
  e_\textrm{d}=\frac{3}{2}\frac{kT_\textrm{d}}{\mu m_\textrm{p}}+\frac{4\sigma}{c\rho_\textrm{d}}T_\textrm{d}^4\;,
\end{equation}
and the corresponding energy for the corona
\begin{equation}
  \label{eq:ec}
  e_\textrm{c}=\frac{3}{2}\frac{k\left(T_\textrm{p,c}+T_\textrm{e,c}\right)}{2\mu m_\textrm{p}}\;.
\end{equation}
Then, with the use of eqns.~(\ref{eq:hydstatdisc}),
(\ref{eq:eosdisc}), (\ref{eq:hydstatcorona2}), (\ref{eq:eoscorona})
and the boundary condition~(\ref{eq:pcpd}) we can solve for
$\rho_\textrm{c}$ and $\rho_\textrm{d}$, the densities in the corona
and disc and the temperature $T_\textrm{d}$ for given $e_\textrm{d}$,
$e_\textrm{c}$, $\Sigma_\textrm{d}=\rho_\textrm{d}H_\textrm{d}$ and
$\Sigma_\textrm{d}=\rho_\textrm{c}H_\textrm{c}$. We arrive at a system
of equations comprising of (\ref{eq:ed}) and
\begin{eqnarray}
%  e_\textrm{d}-\left(\frac{3}{2}\frac{kT_\textrm{d}}{\mu
%      m_\textrm{p}}+\frac{4\sigma}{c\rho_\textrm{d}}T_\textrm{d}^4\right)&=&0\\
  \frac{2}{3}e_\textrm{c}-\Omega^2\frac{\Sigma_\textrm{c}}{\rho_\textrm{c}}\left(\frac{\Sigma_\textrm{d}}{\rho_\textrm{d}}+\frac{1}{2}\frac{\Sigma_\textrm{c}}{\rho_\textrm{c}}\right)&=&0\label{eq:a3}\\
  \rho_\textrm{d}\frac{kT_\textrm{d}}{\mu
    m_\textrm{p}}+\frac{4\sigma}{3c}T_\textrm{d}^4-\left(\frac{2}{3}\rho_\textrm{c}e_\textrm{c}+\frac{1}{2}\Omega^2\frac{\Sigma_\textrm{d}^2}{\rho_\textrm{d}}\right)&=&0\;. \label{eq:a5}
\end{eqnarray}
The electron and proton temperature in the corona follow trivially
from the specific energy of the electrons and protons (see
eq.~\ref{eq:ec}). The variation of $H_\textrm{d}$ and $H_\textrm{c}$
in terms of the internal energies $e_\textrm{d}$ and $e_\textrm{c}$
and surface densities $\Sigma_\textrm{d}$ and $\Sigma_\textrm{c}$ is
given by
\begin{equation}
  \label{eq:hd}
  \frac{dH_\textrm{d}}{H_\textrm{d}}=\left(\frac{8-10\beta_\textrm{P}+3\beta_\textrm{P}^2}{8-7\beta_\textrm{P}}\frac{de_\textrm{d}}{e_\textrm{d}}-\frac{P_\textrm{c}}{2P_\textrm{d}}\frac{H_\textrm{c}}{H_\textrm{c}+H_\textrm{d}}\frac{de_\textrm{c}}{e_\textrm{c}}-\left(1-\frac{P_\textrm{c}}{P_\textrm{d}}+\frac{\beta_\textrm{P}^2+6\beta_\textrm{P}-8}{8-7\beta_\textrm{P}}\right)\frac{d\Sigma_\textrm{d}}{\Sigma_\textrm{d}}-\frac{P_\textrm{c}}{P_\textrm{d}}\frac{d\Sigma_\textrm{c}}{\Sigma_\textrm{c}}\right)/\left(1-\frac{P_\textrm{c}}{P_\textrm{d}}\frac{H_\textrm{c}}{H_\textrm{c}+H_\textrm{d}}+\frac{8-6\beta_\textrm{P}-\beta_\textrm{P}^2}{8-7\beta_\textrm{P}}\right) 
\end{equation}
and
\begin{equation}
  \label{eq:hc}
  \frac{dH_\textrm{c}}{H_\textrm{c}}=\frac{H_\textrm{c}+2H_\textrm{d}}{2\left(H_\textrm{c}+H_\textrm{d}\right)}\frac{de_\textrm{c}}{e_\textrm{c}}-\frac{H_\textrm{d}}{H_\textrm{d}+H_\textrm{c}}\frac{dH_\textrm{d}}{H_\textrm{d}}\;.
\end{equation}

\section{Thermal conduction} \label{sect:conduct}

Thermal conduction occurs in media with a temperature
gradient. Conduction induces an heat flow to cancel this temperature gradient.
In a plasma, electron-proton collision occur typically after the particles
travelled the Debye length. The Debye length $\lambda_\textrm{D}$ is given by
\begin{equation}
  \label{eq:debye}
  \lambda_\textrm{D}=\left(\frac{4\pi \epsilon_0}{e^2}\right)^2 \frac{\left(m u^2\right)^2}{2\pi
    n \ln \Lambda}\;,
\end{equation}
where $\epsilon_0$ is the dielectric constant, $e$ the elementary
charge, $m$ the reduced mass (for electron-proton systems it is
essentially the electron mass), $u$ the relative speed of electron and
proton with respect to each other, $n$ the number density of the
plasma and $\ln \Lambda\approx 20$ the Coulomb logarithm. 

For a thermal plasma, the average Debye length can be found by
integrating over the corresponding isotropic Boltzmann distributions for
electrons and protons, respectively. Thus
\begin{equation}
  \label{eq:debyeth}
  \lambda_\textrm{D}=\left(\frac{4\pi \epsilon_0}{e^2}\right)^2 \frac{9 k_B^2 T_\textrm{e}^2}{2\pi
    n \ln \Lambda}\left(1+\frac{m_\textrm{e}}{m_\textrm{p}}\frac{T_\textrm{p}}{T_\textrm{e}}\right)^2\;,
\end{equation}
where $T_\textrm{e}$ and $T_\textrm{p}$ is the temperature of the
electrons and protons and $k_B$ is the Boltzmann constant. Since both
the electrons and protons have an isotropic velocity distribution, the
effects of alignment and anti-alignment cancel and we do not need to
treat any angle-dependent terms in the velocity $u^2$. For further
reference we define
\begin{equation}
  \label{eq:xi}
  \xi_{ep}=\frac{m_\textrm{e}}{m_\textrm{p}}\frac{T_\textrm{p}}{T_\textrm{e}}\;.
\end{equation}
Note that for a one-temperature plasma ($T_\textrm{e}=T_\textrm{p}$)
the correction factor $1+\xi_{ep}$ is close to unity and the classical
Debye length is reproduced.

The conductive flux of heat then can be described as the thermal
average of $n m u^2 \lambda_\textrm{D} (\partial v/\partial x)$. We find
\begin{eqnarray*}
  q_\textrm{cond}&=&\left<nm_\textrm{e}u^2\lambda_\textrm{D}\frac{\partial u}{\partial
      x}\right>_\textrm{th}\\
  &=&\frac{1}{2}\left<nm_\textrm{e}u\lambda_\textrm{D}\left(\frac{\partial u_\textrm{e}^2}{\partial
      x}+\frac{\partial u_\textrm{p}^2}{\partial
      x}\right)\right>_\textrm{th}\\
  &\approx&
  \frac{1}{2}n\lambda_\textrm{D}\sqrt{\frac{kT_\textrm{e}}{m_\textrm{e}}}\sqrt{1+\xi_{ep}}\left(\frac{\partial \left(kT_\textrm{e}\right)}{\partial x}+\frac{m_\textrm{e}}{m_\textrm{p}}\frac{\partial \left(k T_\textrm{p}\right)}{\partial x}\right)\;,
\end{eqnarray*}
where $u_\textrm{e}$ and $u_\textrm{p}$ are the respective speeds of
the electrons and protons. 
Together with the Debye length (\ref{eq:debyeth}) we find
\begin{equation}
  \label{eq:qcond}
  q_\textrm{cond}=\underbrace{\left(\frac{4\pi
      \epsilon_0}{e^2}\right)^2\frac{9k\left(k_BT_\textrm{e}\right)\frac{5}{2}\left(1+\xi_{ep}\right)^\frac{5}{2}}{4\pi\sqrt{m_\textrm{e}}\ln \Lambda}}_{\kappa_\textrm{cond}}\left(\frac{\partial T_\textrm{e}}{\partial x}+\frac{m_\textrm{e}}{m_\textrm{p}}\frac{\partial T_\textrm{p}}{\partial x}\right)\;.
\end{equation}
Note that we again find for a one-temperature plasma
($T_\textrm{e}=T_\textrm{p}$) the correction factor $1+\xi_{ep}$ is
close to unity and moreover the electron and proton temperature
gradient are the same and hence the proton temperature gradient
contributions to the conductive flux can be neglected. The value of
the parameter $\kappa_\textrm{cond}$ is
\begin{equation}
  \label{eq:ccond}
  \kappa_\textrm{cond}=6.9\cdot 10^{-7}
  T_\textrm{e}^\frac{5}{2}\left(1+\xi_{ep}\right)^\frac{5}{2} \textrm{erg
    s}^{-1} \textrm{cm}^{-1} \textrm{K}^{-1}\;.
\end{equation}
This is somewhat larger than the value quoted in
\citet{1990ApJ...358..375B}. Their value (not containing the
correction factor $1+\xi_{ep}$) is based on
\citet{1984ApJ...281..690D} who use a higher value for the Coulomb
logarithm and furthermore include effects of different ion
species. Our value however is only about 20 per cent larger. Compared
with the other approximations in the model presented in this paper we
consider our value to be sufficiently accurate.

The physical effect presented in this section is similar to
\citet{2007A&A...461..381M}. In both calculations the classical
derivations are extended for a two-temperature plasma. Then the proton
contributions cannot be neglected any more since their plasma speed
can become comparable to the electron plasma speed.

\label{lastpage}

\end{document}